\documentclass[pre, aps, showpacs, superscriptaddress, twocolumn, longbibliography]{revtex4-2}

\usepackage{hyperref}
\hypersetup{
     colorlinks=true,
     linkcolor=Blue,
     filecolor=blue,
     citecolor=blue,      
     urlcolor=Purple,
     }
\usepackage{color}
\usepackage[usenames,dvipsnames]{xcolor}
\usepackage{amsmath,amsthm,amssymb}
\usepackage{graphicx}
\usepackage{float}
\usepackage{epsfig}
\usepackage{bm}% bold math
\usepackage{mathrsfs}
\usepackage{multirow}
\usepackage[all]{xy}
\usepackage{pbox}
\usepackage{verbatim}
\usepackage{braket}
\usepackage{mathtools}
\usepackage{bm}
\usepackage{tikz}
\usepackage{xcolor}
 
\usepackage{mathtools}

\usepackage{mathtools}
\usepackage{tabstackengine}
\usepackage{enumerate}   
\usepackage{wasysym}
\usepackage{xcolor}
\usepackage{nicematrix}
\usepackage{dsfont}
\usepackage{stackengine,scalerel}

\usepackage{enumitem}

\stackMath
\DeclareMathOperator{\Tr}{Tr}

\newcommand{\er}[1]{Eq.~\eqref{#1}}
\newcommand{\ers}[2]{Eqs.~(\ref{#1}-\ref{#2})}
\newcommand{\era}[2]{Eqs.~(\ref{#1}) and (\ref{#2})}

\newcommand{\Er}[1]{Equation~\eqref{#1}}

\newcommand{\beq}{\begin{equation}}
\newcommand{\eeq}{\end{equation}}

\newcommand{\W}{{\mathbb W}}
\newcommand{\WW}{{\mathcal W}}

\renewcommand{\P}{\mathbb P}
\newcommand{\X}{\mathds X}

\begin{document}  

\title{Discrete generative diffusion models without stochastic differential equations: a tensor network approach}

\author{Luke Causer}
\affiliation{School of Physics and Astronomy, University of Nottingham, Nottingham, NG7 2RD, UK}
\affiliation{Centre for the Mathematics and Theoretical Physics of Quantum Non-Equilibrium Systems,
University of Nottingham, Nottingham, NG7 2RD, UK}
\author{Grant M. Rotskoff}
\affiliation{Department of Chemistry, Stanford University, Stanford, California 94305, USA}
\author{Juan P. Garrahan}
\affiliation{School of Physics and Astronomy, University of Nottingham, Nottingham, NG7 2RD, UK}
\affiliation{Centre for the Mathematics and Theoretical Physics of Quantum Non-Equilibrium Systems,
University of Nottingham, Nottingham, NG7 2RD, UK}

\begin{abstract}
    Diffusion models (DMs) are a class of generative machine learning methods that sample a target distribution by transforming samples of a trivial (often Gaussian) distribution using a learned stochastic differential equation.
    In standard DMs, this is done by learning a ``score function'' that reverses the effect of adding diffusive noise to the distribution of interest.
    Here we consider the generalisation of DMs to lattice systems with discrete degrees of freedom, and where noise is added via Markov chain jump dynamics.
    We show how to use tensor networks (TNs) to efficiently define and sample such ``discrete diffusion models'' (DDMs) without explicitly having to solve a stochastic differential equation.
    We show the following:
    (i) by parametrising the data and evolution operators as TNs, the denoising dynamics can be represented exactly;
    (ii) the auto-regressive nature of TNs allows to generate samples efficiently and without bias;
    (iii) for sampling Boltzmann-like distributions, TNs allow to construct an efficient learning scheme that integrates well with Monte Carlo.
    We illustrate this approach to study the equilibrium of two models with non-trivial thermodynamics, the $d=1$ constrained Fredkin chain and the $d=2$ Ising model. 
\end{abstract}

\maketitle

\section{Introduction}\label{sec: introduction}
A central problem in machine learning (ML) is how to train a model to efficiently generate samples from a probability distribution of interest \cite{goodfellow2016deep,mehta2019a-high-bias}. Two typical scenarios are where this target distribution is only known through sampled data, or where relative probabilities are known but the overall normalisation is not \cite{rotskoff2024sampling}. There are many ML strategies to address this problem, a subset of which is based on the idea that a model can be trained to transform a ``noise'' distribution (such as a Gaussian) into a non-trivial distribution of interest over the same domain, in such a way that (easily extractable) noise samples from the first distribution can be transformed into (difficult to generate) samples of the target one. This is the general approach of both ``normalising flows'' \cite{tabak2010density,rezende2015variational,noe2019boltzmann}, and of the so-called {\em diffusion models} \cite{sohl-dickstein2015deep,bahri2020statistical,song2021score-based,yang2023diffusion} that are the focus of this paper.  

Generative diffusion models (DMs) \cite{sohl-dickstein2015deep,bahri2020statistical,song2021score-based,yang2023diffusion}  are a class of machine learning models designed for performing the above transformation by evolving a noisy sample under a stochastic dynamics that undoes the effect of adding noise to the distribution of interest. Given a {\em noising} dynamics such as a simple Brownian process which would convert the distribution of interest into a non-interacting Gaussian, one 
can define a corresponding {\em denoising} dynamics by learning the time-dependent force -- or (Stein) {\em score} \cite{anderson1982reverse-time,song2021score-based} -- 
that needs to be applied (together with the same Brownian noise) to convert over time an initial Gaussian into the original target distribution. While the denoising process can be defined exactly in principle, calculating the score is difficult in practice, and for typical applications where the data is large-dimensional the score is often approximated by a neural network. The noising/denoising can be done directly on the degrees of freedom of the data, or in a lower dimensional feature space that represents the data. The use of DMs has grown to become the method of choice for image generation.

The standard formulation of DMs in terms of Brownian motion and stochastic differential equations presents three main challenges. The first one is the estimation of the denoising SDE via the score \cite{song2021score-based}, which has to be learned as a function over the whole domain of the target probability from sparse and high dimensional training data. The second one is how to resolve the ``mismatch in time'' \cite{de-bortoli2021diffusion}: under Brownian dynamics the mapping from the target distribution to a noisy Gaussian happens only asymptotically, while in practice the denoising process is run over finite times, thus incurring in a reconstruction error. The third challenge is how to precisely estimate the likelihood of generated configurations from the learned score. 

In this paper we show how to address these three challenges by integrating DMs with tensor networks (TNs) for the case where the system of interest is defined on a lattice with discrete local degrees of freedom. Initially developed for the study of quantum many-body systems, TNs~\cite{verstraete2008matrix,schollwock2011the-density-matrix,orus2014a-practical,silvi2019the-tensor,okunishi2022developments,banuls2023tensor} are an efficient parametrisation of many-body states and operators in terms of graphs of local tensors encoding locality properties of the systems under study. 
TN methods are being increasingly applied in the context of classical stochastic dynamics, in particular to compute statistical properties of dynamical trajectories and for the study of rare events, see e.g.\ Refs.~
\cite{gorissen2009density-matrix,gorissen2012current,banuls2019using,helms2019dynamical,helms2020dynamical,causer2020dynamics,causer2022finite,strand2022using,causer2023optimal}. 
Here we show how to use TNs for ``discrete diffusion models'' (DDMs), where the underlying dynamics is not Brownian but that of Markov jump processes, thus bypassing the need to solve a stochastic differential equations for obtaining the denoising process. 
Discrete diffusion models~\cite{austin2021structured, campbell2022a-continuous} have recently attracted significant interest for applications in natural language processing~\cite{austin2021structured}, protein sequence modelling~\cite{nisonoff2024unlocking, gruver2023protein, morehead2023towards}, but have been implemented using the conventional toolkit for continuous-time Markov chains. 
We show how to efficiently define, train and run DDMs by parametrising both the probability vectors and the evolution operators as TNs, so that the denoising dynamics can be represented exactly. Furthermore, the auto-regressive property of TNs makes the generation of samples efficient and free of bias. We focus on the problem of sampling Boltzmann-like distributions \cite{rotskoff2024sampling}, and show that DDMs with TNs allow to construct an efficient scheme that integrates well with Monte Carlo sampling. For illustration, we use this DDM enhanced Monte Carlo scheme to study the equilibrium properties of two models with non-trivial thermodynamics, the one-dimensional constrained Fredkin spin chain \cite{causer2022slow} and the two-dimensional Ising model \cite{chandler1987introduction}.

\section{Tensor networks}\label{sec: mps}
In what follows we consider for simplicity lattice systems of $N$ sites, with each site $j = 1, \dots, N$ hosting a binary variable $\sigma_{j}$, which we will refer to as ``spin'', with $\sigma_{j} \in \{-1, +1\}$ (or $\sigma_{j} \in \{0, 1\}$ depending on the specific model). We define a {\em target} probability distribution $P$ over the system which we write as the vector
\beq
    \ket{P} = \sum_{\bm \sigma} P({\bm \sigma}) \ket{\bm \sigma},
    \label{P}
\eeq
where the $\ket{\bm \sigma} = \ket{\sigma_{1}, \dots, \sigma_{N}}$ are configuration basis states, $P({\bm \sigma})$ is the probability of configuration ${\bm \sigma}$, and $\sum_{\bm \sigma} P({\bm \sigma}) = 1$.
The class of problems we will focus on are those where one knows the functional form of this target probability up to an overall constant, as for example in the general case of Boltzmann sampling where
\begin{equation}
    P({\bm \sigma}) = \frac{e^{-\beta E(\bm \sigma)}}{Z_\beta}
    \label{eq:Boltzmann}
\end{equation}
for a known energy function $E({\bm \sigma})$, but where the partition function 
\begin{equation}
    Z_\beta \equiv \sum_{\bm \sigma} e^{-\beta E(\bm \sigma)}
    \label{Z}
\end{equation}
is unknown and difficult/impossible to calculate explicitly. Our aim is to find an efficient method for sampling \er{P} using Markov Chain Monte Carlo (MCMC). We will achieve this by defining an approximate distribution $\ket{P_{{\bm \theta}}}$, where ${\bm \theta}$ are the parameters that define a convenient variational class for probability vectors, so that when properly trained, $\ket{P_{{\bm \theta}}}$ allows $\ket{P}$ to be sampled efficiently. The training of 
$\ket{P_{{\bm \theta}}}$ will be done by evolving probability vectors via a combination of {\em noising} and {\em denoising} dynamical protocols common to DMs, as explained in detail in Sec.~\ref{sec: denoising}.

The implementation of the noising and denoising protocols will require the following: (i) an efficient representation of probability vectors for large system sizes $N$ (i.e., the variational class represented by ${\bm \theta}$), and (ii) the ability to time evolve these vectors under continuous-time Markov dynamics.
One framework that satisfies both of these conditions is that of tensor networks (TNs); for reviews see e.g. Refs.~\cite{orus2019tensor,banuls2023tensor}.
TNs are an efficient decomposition of high-dimensional objects (such as vectors of many-body systems) into contractions of smaller tensors, which allows calculations to be done in a tractable way. 
In this paper, we will focus on {\em matrix product states} (MPS) to represent vectors, and {\em matrix product operators} (MPOs) to represent the operators that act on them; for reviews see e.g. Refs.~\cite{schollwock2011the-density-matrix,cirac2021matrix}.

While originally designed to study the ground state \cite{white1992density, ostlund1995thermodynamic} 
and dynamics \cite{vidal2004efficient, daley2004time-dependent} 
of one-dimensional quantum many-body systems, MPS have also proven useful in capturing the properties of classical models in statistical mechanics 
\cite{honecker1997matrix-product, hieida1998application,kemper2002stochastic, ueda2005snapshot,frias-perez2023collective} and studying the non-equilibrium dynamics of classical stochastic dynamics
\cite{gorissen2009density-matrix, gorissen2012current, gorissen2012exact,banuls2019using, helms2019dynamical, causer2020dynamics, causer2021optimal,causer2022finite, gu2022tensor-network, strand2022using, causer2022slow, garrahan2022topological, nicholson2023quantifying, merbis2023efficient}.
Recent works have also demonstrated how TNs can be used to efficiently study rare-events in classical stochastic systems through the realisation of the optimal sampling (so-called ``Doob'') dynamics \cite{causer2021optimal, causer2022finite, causer2023optimal}. This latter approach will be used here to implement the denoising protocol defined in Sec.~\ref{sec: denoising}.

\subsection{Probability distributions as matrix product states}
In general, any vector defined on the space of $N$ binary variables can be written as
\beq
    \ket{\psi} = \sum_{\bm \sigma} \psi(\bm \sigma) \ket{{\bm \sigma}}.
    \label{eq:psi}
\eeq
We then write $\psi({\bm \sigma}) = \psi(\sigma_{1} ,\dots, \sigma_{N})$ as a product of matrices 
\beq
    \psi({\bm \sigma}) = \Tr \left[M^{(1)}_{\sigma_{1}} \cdots M^{(N)}_{\sigma_{N}}\right],
    \label{eq:MPS}
\eeq
where each $M^{(j)}_{\sigma_{j}}$ is a $D\times D$ matrix, with the virtual {\em bond dimension} $D$.
When $\sigma_{j}$ is unspecified, $M^{(j)}_{\sigma_{j}} := M^{(j)}$ is a rank-3 tensor with dimensions $D \times D \times 2$.
Equation \eqref{eq:MPS} is often referred to as an MPS, and can be thought of as an efficient representation of a large one-dimensional vector space.
The bond dimension controls the extent of correlations that can be described by the MPS:
the MPS is known to obey an {\rm area law} \cite{verstraete2006matrix,hastings2007an-area}, meaning that it can efficiently capture states that have finite correlation lengths, such as the ground states of one-dimensional quantum many-body systems.
In quantum many-body systems, the von Neumann entanglement entropy (a quantum analogue to mutual information) between two partitions of the state space that can be described by an MPS with bond dimension $D$ is bound by $S_{E} \leq 2\log D$.

\begin{figure}[t]
    \centering
    \includegraphics[width=\linewidth]{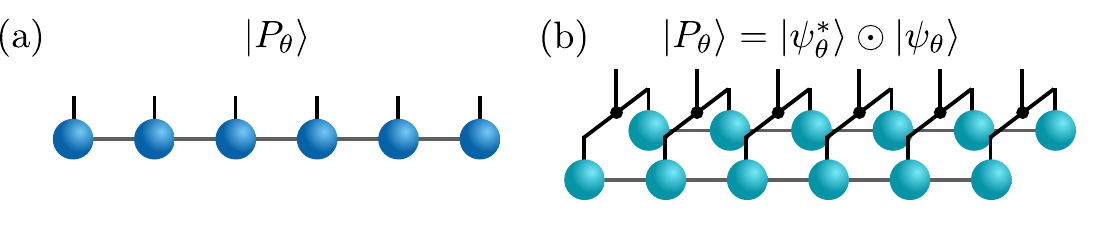}
    \caption{\textbf{Tensor networks.} 
    Diagrammatic representation of tensor networks.
    (a) The probability vector $\ket{P_{{\bm \theta}}}$ as an MPS.
    Each vertex is a rank-3 tensor for lattice $j$.
    The grey edges connecting neighbouring vertices represent a contraction over the virtual dimension of size $D$.
    The open black edges represent the physical dimensions.
    (b) The probability vector with a positivity ansatz, $\ket{P_{\bm \theta}} = \ket{\psi^{*}_{\bm \theta}} \odot \ket{\psi_{\bm \theta}}$.
    The physical dimensions of the two MPS are contracted with a three-point delta function.
    }
    \label{fig: mps}
\end{figure}

It is often convenient to work with TNs in a diagrammatic notation, where shapes represent tensors, and edges emerging from the shapes represent a dimension of the tensor.
Legs that connect two shapes represent the contraction over tensors.
We show the diagrammatic representation for an MPS in Fig.~\ref{fig: mps}(a):
each binary variable (or lattice site) has its own tensor, which is contracted with the tensors of neighbouring lattice sites through the virtual bond dimension.
The open black edges represent the physical dimensions of the system (i.e. each degree of freedom).

While MPS provide an efficient way to classically represent or estimate a large class of vectors for an exponentially growing state space, there is one challenge to overcome for describing probability vectors, $\ket{P_{\bm \theta}}$. 
That is, each element of the probability vector must be a non-negative real number, $P_{\bm \theta}({\bm \sigma}) \geq 0$.
This is hard to enforce (or to even verify) for an arbitrary MPS. 
One way to overcome this difficulty is to use a {\em positivity ansatz} (sometimes referred to as a {\em Born machine}) \cite{han2018unsupervised,lin2023distributive},
\beq
    P_{\bm \theta}({\bm \sigma}) = \psi^{*}_{\bm \theta}({\bm \sigma}) \psi_{\bm \theta}({\bm \sigma}),
    \label{eq:pa}
\eeq
or as a vector, $\ket{P_{\bm \theta}} = \ket{\psi^{*}_{\bm \theta}} \odot \ket{\psi_{\bm \theta}}$, where $\odot$ denotes the {\em Hadamard product} of two vectors.
This is shown in Fig.~\ref{fig: mps}(b), where each dot with three legs is the three-point delta function.
In principle, it is possible to represent this as a single MPS (noting that if the bond dimension of $\ket{\psi}$ is $D$, then the bond dimension of the probability vector will be $D' \leq D^{2}$).
However, due to our choice of local noising dynamics (see below), it will be convenient to work with the positivity ansatz \eqref{eq:pa} directly. 

\section{Generative Discrete Diffusion Protocols with Tensor Networks}\label{sec: denoising}
Given the target distribution $P$ of the form \eqref{eq:Boltzmann}, our aim is to learn an approximate state $\ket{P_{{\bm \theta}}}$ which allows $P$ to be sampled efficiently, where the trainable parameters ${\bm \theta}$ of $\ket{P_{{\bm \theta}}}$ indicate the tensors of the MPS that defines it, \ers{eq:psi}{eq:pa}. Our method is based on discrete diffusion models (DDMs) \cite{sohl-dickstein2015deep}.
These are generative models capable of learning the underlying distributions of complex datasets in order to sample them efficiently. They are defined in terms of two stochastic processes. The first one is {\em noising} (sometimes called the ``forward'' process), where a stochastic dynamical evolution progressively corrupts a dataset, eventually mapping it onto a distribution of noise which can be easily sampled (in contrast to the underlying distribution of the initial dataset). The noising step is easy to define, for example for continuous degrees of freedom it can be realised with a simple Brownian process (i.e.\ simple diffusion), which asymptotically converts any initial distribution into a Gaussian. The second step is to define a {\em denoising} (or ``reverse'' or ``backward'') process, whose aim is to undo the effect of the noise, by mapping the final distribution of the noising dynamics to the initial one, through all of the same intermediate distributions. 
To do this it is necessary to implement a bias in the transition probabilities of the denoising process, which for the case of diffusions is known as the (Stein) ``score'' \cite{song2020generative, song2021score-based}, but also generalises to the discrete case \cite{ho2020denoising}. Defining the denoising dynamics is the hard part of these methods. 

For a discrete state space, the scoring approach can be generalised to a continuous-time Markov process which learns the time-dependent transition rates for the denoising process \cite{sun2023score-based}.
This approach effectively learns the {\em generalised Doob transform} \cite{chetrite2015nonequilibrium} that 
reverses the noising dynamics while maintaining stochasticity (i.e., probability conservation). In Refs.~\cite{causer2021optimal, causer2022finite, causer2023optimal}, it was shown how TNs can be used it efficiently approximate the Doob dynamics for sampling dynamical large deviations. Here, we generalise this same approach for DDMs. One of the key benefits of TNs is that they can efficiently model the dynamics at the level of the master equation, which is equivalent to evolving the entire ensemble of trajectories, as opposed to simply sampling them. 
As we explain below, this can be very beneficial when combined with MCMC to sample the distribution of interest \cite{hunt-smith2024accelerating}.

\begin{figure}[t]
    \centering
    \includegraphics[width=0.6\linewidth]{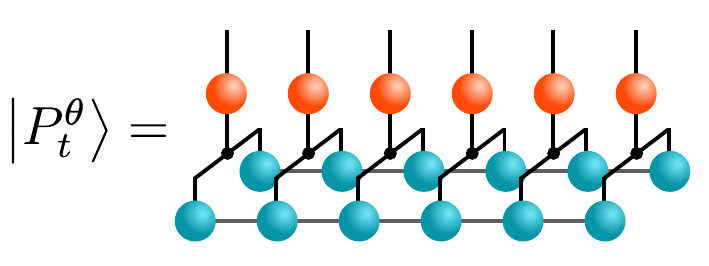}
    \caption{\textbf{The noising protocol as a TN.}
    The distribution $\ket{P^{\bm \theta}_t} = \WW_{t \leftarrow 0} \ket{P_{\bm \theta}}$ can be efficiently described by a TN.
    The blue and black spheres are the tensors for the probability $\ket{P_{{\bm \theta}}}$, see Fig.~\ref{fig: mps}(b).
    The orange spheres are the tensors for the evolution operator $\WW_{t \leftarrow 0}$, see \er{eq:noising_protocol}.
    }
    \label{fig: noising_mps}
\end{figure}

\subsection{Noising protocol}
We define the noising dynamics as a continuous-time Markov process with a (time independent) Markov generator $\W$ and some initial distribution $\ket{P^{{\bm \theta}}_{t=0}}$.
That is, the time-dependent probability distribution $\ket{P^{\bm \theta}_t}$ at time $t>0$, whose components are $P^{\bm \theta}_t({\bm \sigma})$, evolves under the master equation
\beq
    \partial_{t} \ket{P^{\bm \theta}_t} = \W \ket{P^{\bm \theta}_t}
    \label{ME}
\eeq
for times $0 \leq t \leq T$, where $T$ is the maximum noising time and $\ket{P^{\bm \theta}_{t=0}} = \ket{P_{\bm \theta}}$ is the MPS we aim to learn. 
Since $\W$ is a Markov generator, its general form is 
\beq   
    \W =
        \sum_{{\bm \sigma} \neq {\bm \sigma}'}
        W({{\bm \sigma}' \leftarrow {\bm \sigma}})
        \ket{{\bm \sigma}'} \bra{{\bm \sigma}}
        -
        \sum_{{\bm \sigma}}
        R({{\bm \sigma}})
        \ket{{\bm \sigma}} \bra{{\bm \sigma}} ,
    \label{eq:WR}
\eeq
where $W({{\bm \sigma}' \leftarrow {\bm \sigma}})$ are the transition rates between configurations and $R({{\bm \sigma}})$ the escape rates. It follows that the time-evolved distribution at time $t' > t$ can be calculated by formally integrating \eqref{ME} 
\beq
    \ket{P^{{\bm \theta}}_{t_2}} = e^{(t_2 - t_1)\W} \ket{P^{{\bm \theta}}_{t_1}} = \WW_{t_2 \leftarrow t_1} \ket{P^{{\bm \theta}}_{t_1}},
    \label{eq:Ptheta}
\eeq
with the propagator
\beq
    \WW_{t_2 \leftarrow t_1} \equiv e^{(t_2-t_1) \W} ,
    \label{WW}
\eeq
depending only on the time difference due to the time-homogeneity of $\W$. 

We choose $\W$ such that its stationary state is a {\em noise distribution} that is easy to sample. Specifically, we choose $\W$ to be {\em bistochastic} 
\beq
    \bra{-} \W = 0, \;\;\; 
    \W \ket{-} = 0 ,
\eeq
meaning that its stationary state coincides with the {\em uniform distribution} (or ``flat state'')
\beq
    \ket{-} = 2^{-N}\sum_{\bm \sigma} \ket{\bm \sigma} .
    \label{eq:fs}
\eeq
This means that at long times the dynamics generated by $\W$ will converge to \eqref{eq:fs}. For the spin problems we will consider the flat state is the product Bernoulli measure 
\beq
    \ket{-} = 2^{-N} \left( \ket{+1} + \ket{-1} \right)^{\otimes N} ,
    \label{eq:Bernoulli}
\eeq
which is straightforward to sample by flipping $N$ fair coins.

The simplest bistochastic noising dynamics we can choose for spin systems is that of {\em non-interacting} single-spin flips generated by 
\beq
    \W = \sum_{j=1}^{N} (\X_{j} - \mathds{1}),
    \label{eq:W}
\eeq
where $\X_{j}\ket{\sigma_{j}} = \ket{-\sigma_{j}}$ is the operator which flips the spin $j$, and $\mathds{1}$ is the identity matrix. The generator \eqref{eq:W} has transition rates $W({{\bm \sigma}' \leftarrow {\bm \sigma}})=1$ between all ${\bm \sigma}$ and ${\bm \sigma}'$ differing by a single spin flip (zero otherwise), and escape rates $R({{\bm \sigma}})=N$ for all ${\bm \sigma}$. 

\begin{figure}[t]
    \centering
    \includegraphics[width=0.9\linewidth]{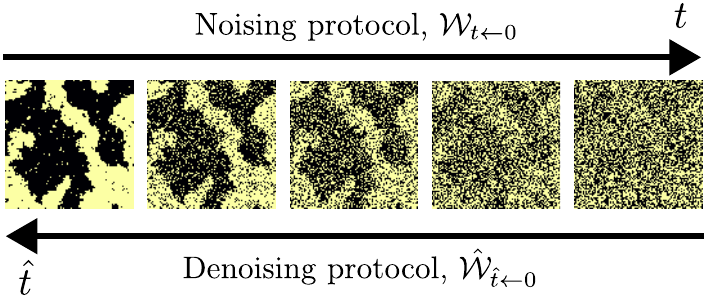}
    \caption{\textbf{Noising and denoising protocols}.
    The noising protocol is a continuous-time Markov dynamics, $\WW_{t \leftarrow 0}$, which progressively evolves a distribution onto the uniform distribution, $\ket{-}$.
    The denoising protocol is a time-inhomogeneous Markov dynamics, $\hat{\WW}_{\hat{t} \leftarrow 0}$, which reverses the noising process.
    }
    \label{fig: noise-denoise}
\end{figure}

The simplicity of \er{eq:W} allows us to calculate the evolution operator \eqref{WW} exactly:
\beq
    \WW_{t_2 \leftarrow t_1} = \bigotimes_{j=1}^{N} \left(\frac{1 + e^{-(t_2-t_1)}}{2}\mathds{1} + \frac{1 - e^{-(t_2-t_1)}}{2} \X_j \right).
    \label{eq:noising_protocol}
\eeq
Since the noising dynamics is non-interacting, the evolution operator \eqref{eq:noising_protocol} is the tensor product of local operators, and thus can be efficiently implemented as an MPO with bond dimension $\chi_{O} = 1$.
This means that if the initial state $\ket{P_{{\bm 
\theta}}}$ is an MPS then the evolved state $\ket{P^{\bm \theta}_t} = \WW_{t \leftarrow 0} \ket{P_{\bm \theta}}$ can be represented efficiently as a TN, see Fig.~\ref{fig: noising_mps}. An example of the effect of noising is given in Fig.~\ref{fig: noise-denoise}.

\subsection{Denoising protocol}

Given the noising dynamics above, we can define an associated denoising protocol that inverts its action \cite{sun2023score-based}, as sketched in Fig.~\ref{fig: noise-denoise}. 
We denote the denoising evolution operator $\hat{\WW}^{\bm \theta}_{\hat{t}_{2} \leftarrow \hat{t}_{1}}$ and we label by $\hat{t}$ the time when running this dynamics. The denoising dynamics is defined in terms of a discrete version of the ``score'' \cite{sun2023score-based}. If the noising dynamics goes from $t=0$ to $t=T$, then the denoising evolution is \cite{sun2023score-based}
\begin{equation}
    \hat{\WW}^{\bm \theta}_{\hat{t}_{2} \leftarrow \hat{t}_{1}} = 
    \P^{\bm \theta}_{T-\hat{t}_{2}}
    \WW^{\rm T}_{T-\hat{t}_{1} \leftarrow T-\hat{t}_{2}}
    (
    \P^{\bm \theta}_{T-\hat{t}_{1}}
    )^{-1} ,
    \label{eq:tWW}
\end{equation}
where $\P^{\bm \theta}_{t}$ is the diagonal operator 
$\P^{\bm \theta}_{t} = \sum_{\bm \sigma} P^{\bm \theta}_{t}({\bm \sigma}) \ket{\bm \sigma} \bra{\bm \sigma}$, and $(\P^{\bm \theta}_{t})^{-1}$ its inverse. Note that denoising dynamics \eqref{eq:tWW} is explicitly time-dependent, with $\hat{t}$ running between $\hat{t}=0$ and $\hat{t}=T$, 
and also depends on the parameters of the initial distribution $P_{\bm \theta}$ due to presence of the noise-evolved states in \eqref{eq:tWW}. Furthermore, by differentiating the relation 
\beq
    \hat{\WW}^{\bm \theta}_{\hat{t}_{2} \leftarrow \hat{t}_{1}} 
        = 
        \exp{\int_{\hat{t}_{1}}^{\hat{t}_{2}} \hat{\W}^{\bm \theta}_{\hat{t}} \, d\hat{t} }
    \label{eq:tWWtW}
\eeq
with respect to time, we obtain the time-dependent generator of the denoising dynamics corresponding to \eqref{eq:tWW}
\beq
    \hat{\W}^{\bm \theta}_{\hat{t}}
        = 
        \P^{\bm \theta}_{T-\hat{t}}
        \W^{\rm T}
        (
        \P^{\bm \theta}_{T-\hat{t}}
        )^{-1} 
        +
        \frac{d}{d \hat{t}}
        \ln 
        \P^{\bm \theta}_{T-\hat{t}} ,
    \label{eq:tW}
\eeq
so that the denoising transition and escape rates are 
\begin{align}
    W^{\bm \theta}_{\hat{t}}({\bm \sigma}' \leftarrow {\bm \sigma})
        &= 
        \frac{P^{\bm \theta}_{T-\hat{t}}({\bm \sigma}')}
        {P^{\bm \theta}_{T-\hat{t}}({\bm \sigma})}
        W({\bm \sigma} \leftarrow {\bm \sigma}'),
    \label{eq:tWmat}
    \\
    R^{\bm \theta}_{\hat{t}}({\bm \sigma})
        &=
        R({{\bm \sigma}}) - 
        \frac{d}{d \hat{t}} 
        \ln P^{\bm \theta}_{T-\hat{t}}({\bm \sigma}).
    \label{eq:tR}
\end{align}

The form of the denoising generator \eqref{eq:tW} is that of a ``generalised Doob transform'' \cite{chetrite2015nonequilibrium}, often encountered in the context of dynamical large deviations \cite{touchette2009the-large,garrahan2018aspects,jack2020ergodicity}. This connection is as follows. Consider a {\em tilting} \cite{touchette2009the-large,garrahan2018aspects,jack2020ergodicity} of the generator \eqref{eq:W}, $\W \to \W_\lambda$, where in $\W_\lambda$ the transition probabilities change to 
$W_\lambda({{\bm \sigma}' \leftarrow {\bm \sigma}}) \equiv 
W({{\bm \sigma}' \leftarrow {\bm \sigma}}) e^{\lambda 
\ln [W({{\bm \sigma}' \to {\bm \sigma}})/W({{\bm \sigma}' \leftarrow {\bm \sigma}})]}$, while the escape rates remain the same. The tilted generator $\W_\lambda$ corresponds to an exponential reweighting of the probabilities of trajectories of the noising dynamics, which for $\lambda=1$ is equivalent to transposing the evolution operator of the noising dynamics, $e^{t \W_{\lambda = 1}} = \WW_{t \leftarrow 0}^{\rm T}$. While the operator $\W_\lambda$ is not stochastic (i.e., $\bra{-} \W_\lambda \neq 0$ in general), it can be brought to a stochastic form through the gauge transformation \cite{garrahan2016classical} that defines the denoising generator \eqref{eq:tW}.

\begin{figure}[t]
    \centering
    \includegraphics[width=\linewidth]{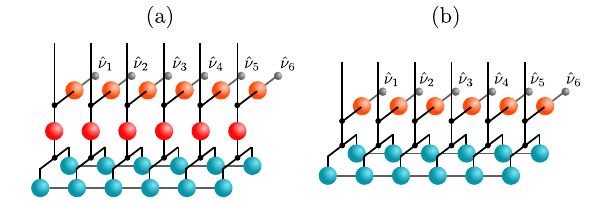}
    \caption{\textbf{Denoising protocol as a TN.}
    (a) Graphical representation of \er{eq:Q_tau_cond}: the state $\ket{\hat{P}^{\bm \theta}_{\hat{t} | \hat{\bm{\nu}}}}$ is an MPS obtained from propagating the initial $\ket{P_{\bm \theta}} = \ket{\psi^{*}} \odot \ket{\psi}$, where the blue spheres represent $\psi$, with the noising evolution operator $\WW_{T-\hat{t}\leftarrow 0}$, represented by the red spheres, to obtain $\ket{P^{{\bm \theta}}_{T-\hat{t}}}$. The small grey spheres indicate the initial state $\hat{{\bm \nu}} = (\hat{\nu}_{1} ,\dots, \hat{\nu}_{N})$ for the denoising, which is acted upon by $\WW^{\rm T}_{0 \to \hat{t}}$ (orange spheres, and where the black circles indicate delta tensors), and multiplied element-wise to $\ket{P^{{\bm \theta}}_{T-\hat{t}}}$. Rescaling by the overall factor $1/\braket{\hat{\bm{\nu}}|P^{\bm \theta}_{T}}$ (not shown) gives $\ket{\hat{P}^{\bm \theta}_{\hat{t} | \hat{\bm{\nu}}}}$. (b) Graphical representation when $\hat{t}=T$, see \er{eq:Qt_update}: in this case there is no propagation of $\ket{P_{\bm \theta}}$. 
    }
    \label{fig: denoising_sampling}
\end{figure}

While the evolution under the noising dynamics, \era{eq:Ptheta}{eq:noising_protocol}, when starting from an MPS is easy to compute as a TN, the evolution under the denoising dynamics,
\beq
    \ket{\hat{P}^{\bm \theta}_{\hat{t}}} = 
        \hat{\WW}^{\bm \theta}_{\hat{t} \leftarrow 0} 
        \ket{\hat{P}_0} ,
    \label{eq:tP}
\eeq
in general might be difficult, even if $\ket{\hat{P}_0}$ is an MPS. This is because $(\P^{\bm \theta}_{t})^{-1}$ cannot be efficiently represented as a TN. Consider instead starting the denoising dynamics from one configuration $\hat{\bm{\nu}}$. After denoising for time $\hat{t}$ we obtain a probability $\hat{P}^{\bm \theta}_{\hat{t} | \hat{\bm{\nu}}}$ conditioned on the initial $\hat{\bm{\nu}}$
\begin{align}
    \ket{\hat{P}^{\bm \theta}_{\hat{t} | \hat{\bm{\nu}}}} 
    &= 
    \P^{\bm \theta}_{T-\hat{t}}
    \WW^{\rm T}_{\hat{t} \leftarrow 0}
    \ket{\hat{\bm{\nu}}} 
    \frac{1}{P^{\bm \theta}_{T}(\hat{\bm{\nu}})}
    \nonumber    \\
    & = 
    \ket{P^{{\bm \theta}}_{T-\hat{t}}}
    \odot
    \WW^{\rm T}_{\hat{t} \leftarrow 0}
    \ket{\hat{\bm{\nu}}}
    \frac{1}{P^{\bm \theta}_{T}(\hat{\bm{\nu}})} .
    \label{eq:Q_tau_cond}
\end{align}
The factor $1/P^{\bm \theta}_{T}(\hat{\bm{\nu}})$ is easily obtained using the noising protocol described above 
since $\ket{P^{\bm \theta}_{T}}$ is an MPS and $P^{\bm \theta}_{T}(\hat{\bm{\nu}}) = \braket{\hat{\bm{\nu}} | P^{\bm \theta}_{T}}$ can be efficiently extracted from it. The other factors are an MPS and an MPO, so we can then calculate \er{eq:Q_tau_cond} efficiently as a TN. This is shown graphically in Fig.~\ref{fig: denoising_sampling}(a). 

A key benefit of formulating the problem in this way is that given some noise sample $\hat{\bm{\nu}}$ we can exactly calculate the marginal distribution at time $\hat{t} = T$, where the $\ket{P^{{\bm \theta}}_{T-\hat{t}}}$ factor in \eqref{eq:Q_tau_cond} reduces to the initial MPS, 
\beq
    \ket{\hat{P}^{\bm \theta}_{T | \hat{\bm{\nu}}}} 
    = 
    \ket{P_{{\bm \theta}}}
    \odot
    \WW^{\rm T}_{T \leftarrow 0}
    \ket{\hat{\bm{\nu}}} 
    \frac{1}{P^{\bm \theta}_{T}(\hat{\bm{\nu}})} ,
\label{eq:Qt_update}
\eeq
simplifying the TN as shown in Fig.~\ref{fig: denoising_sampling}(b). 
\Er{eq:Qt_update} represents the final state of the denoising dynamics over all possible trajectories that start at fixed $\hat{\bm{\nu}}$. This allows us to obtain denoised configurations efficiently from the by directly sampling the MPS \eqref{eq:Qt_update}. [If we are interested in sampling the whole denoising trajectory, and not only the final denoised state, then we can run explicitly the denoising dynamics using \ers{eq:tW}{eq:tR}.]

If the initial state $\hat{\bm{\nu}}$ is sampled from an initial denoising probability $\ket{\hat{P}_0}$ then the corresponding final denoising probability vector reads
\begin{align}
    \ket{\hat{P}^{\bm \theta}_{T}}
    &=
    \sum_{\hat{{\bm \nu}}} 
    \ket{\hat{P}^{\bm \theta}_{T|\hat{{\bm \nu}}}}
    \braket{\hat{{\bm \nu}} | \hat{P}_0} 
    \nonumber \\
    &=
    \sum_{\hat{{\bm \nu}}} 
    \ket{P_{{\bm \theta}}}
    \odot
    \WW^{\rm T}_{T \leftarrow 0}
    \ket{\hat{\bm{\nu}}}
    \frac{\hat{P}_0(\hat{\bm{\nu}})}{P^{\bm \theta}_{T}(\hat{\bm{\nu}})} .
    \label{eq:QT}
\end{align}
We note two things. Firstly, by definition of the denoising dynamics, see \er{eq:tWW}, if the initial samples for denoising come from the final noising state, 
$\hat{P}_0 = P^{\bm \theta}_{T}$, then 
sampling from \er{eq:QT} is the same as sampling from the initial $P_{\bm \theta}$. However, as is standard in DMs, one often wishes to initiate the denoising from purely noise samples (which are easy to generate), corresponding in our case to $\ket{\hat{P}_0} = \ket{-}$, which only coincides with the final noising state for $T \to \infty$.  Therefore denoising for finite $T$ implies a ``mismatch'', and the endpoint of the denoising dynamics does not strictly coincide with $P_{\bm \theta}$ (see e.g.\ \cite{de-bortoli2021diffusion} for discussions on this point). Secondly, in contrast to \er{eq:Qt_update}, the state \eqref{eq:QT} is not an MPS, which makes extracting the probability for specific configurations, $\hat{P}^{\bm \theta}_T({\bm \sigma}) = \braket{{\bm \sigma} | \hat{P}^{\bm \theta}_T}$, difficult to compute. These two issues will inform the sampling strategies we define below.

\section{Noising-denoising as generative updates for MCMC}
\label{sec: protocols}
We wish to sample a distribution $P$ of the form \er{eq:Boltzmann} by means of MCMC using our noising and denoising protocols as a proxy. We can do this by 
generating the proposed Monte Carlo moves 
using noising-denoising with initial distribution $P_{{\bm \theta}}$ in MPS form, which in general is only an approximation to the target $P$. The aim will be to eventually learn the parameters ${\bm \theta}$ to optimise convergence of the MCMC dynamics. 

\begin{figure}[t]
    \centering
    \includegraphics[width=0.9\linewidth]{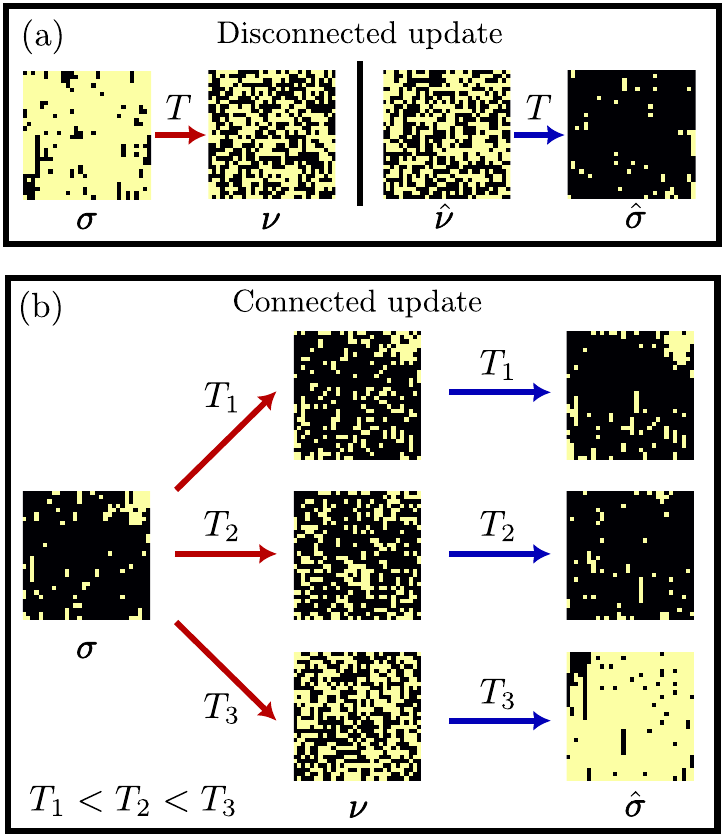}
    \caption{\textbf{MCMC updates with DDM generated proposals}.
    (a) Disconnected update: We sample $P({\bm \sigma})$ by sampling the joint $P_{0,T}({\bm \sigma},{\bm \nu})$ and contracting as the acceptance probability \er{eq:Metropolis2} can be computed efficiently with TNs while the naive \er{eq:Metropolis} cannot. A proposed new pair $(\hat{\bm \sigma},\hat{\bm \nu})$ is obtained by drawing $\hat{\bm \nu}$ from the flat distribution and applying the denoising protocol for time $T$ to generate $\hat{\bm \sigma}$. The new pair is accepted with probability \er{eq:acceptance_correlated}. The generated $\hat{\bm \sigma}$ are always uncorrelated from the current state ${\bm \sigma}$ since the starting configuration $\hat{\bm \nu}$ of the denoising step is completely independent of final configuration ${\bm \nu}$ of the noising step. 
    ---
    (b) Connected update:
    We sample $P({\bm \sigma})$ directly. Starting from the current configuration ${\bm \sigma}$ we denoise it for time $T$, producing a corrupted configuration ${\bm \nu}$.
    We then run denoising starting from ${\bm \nu}$ also for time $T$ to generate $\hat{\bm \sigma}$. This proposal is accepted with probability \eqref{eq:acceptance_correlated} 
    which can be efficiently computed with TNs. 
    The proposed $\hat{\bm \sigma}$ is correlated with the current ${\bm \sigma}$ through ${\bm \nu}$. The 
    degree of correlation is controlled by $T$, with shorter $T$ corresponding to stronger correlation. 
    (We show configurations sampled using an MPS for a 2D Ising model of size $N = 30 \times 30$ with open boundary conditions and at inverse temperature $\beta = 1.1\beta_{c}$, where $\beta_{c}$.)
    }
    \label{fig: protocols}
\end{figure}

We will consider two different strategies:
(i) a {\em disconnected update}, which proposes a new configuration from a random sample of noise, and is therefore entirely uncorrelated from the previous sample, see Fig.~\ref{fig: protocols}(a) 
for an illustration; 
and (ii) a {\em connected update}, which uses a noising and denoising cycle to propose an update which is correlated to the previous sample, as illustrated in Fig.~\ref{fig: protocols}(b).

\subsection{Disconnected update}
\label{sec: disconnected}

Suppose at a given iteration in the MCMC the last accepted  configuration is ${\bm \sigma}$. The next MCMC iteration requires a proposed new configuration $\hat{{\bm \sigma}}$ to attempt to move to. A way to integrate our DDM with Monte Carlo is to generate this proposal by running the denoising dynamics starting from a random configuration $\hat{{\bm \nu}}$, i.e., taking as $\ket{\hat{P}_0}$ the flat state, $\ket{\hat{P}_0} = \ket{-}$. The proposed configuration $\hat{{\bm \sigma}}$ is obtained by denoising for time $\hat{t}=T$ from \eqref{eq:Qt_update}
\footnote{As explained after \er{eq:QT}, for finite $T$ the generated $\hat{{\bm \sigma}}$ will not be equivalent to samples from the initial $P_{\bm \theta}$ which defines the dynamics. But since we only require proposals, to be accepted or rejected according to a Metropolis criterion, this ``mismatch'' is not an issue in our case.}.

Once a new configuration $\hat{{\bm \sigma}}$ is proposed, it is accepted or rejected via a Metropolis test with the usual acceptance probability
\beq
    \mathcal{A}(\hat{{\bm \sigma}} | {\bm \sigma}) 
    = 
    \min 
    \left[ 
        1, \frac{P(\hat{{\bm \sigma}})}{P({\bm \sigma})} \frac{\hat{P}^{\bm \theta}_T({\bm \sigma})}{\hat{P}^{\bm \theta}_T(\hat{{\bm \sigma}})} 
    \right] ,
    \label{eq:Metropolis}
\eeq
dependent on the probability $P({\bm \sigma})$ one wishes to sample, and the probability of proposals $\hat{P}^{\bm \theta}_T({\bm \sigma}) = \braket{{\bm \sigma} | \hat{P}^{\bm \theta}_T} = 2^{-N} \sum_{\hat{{\bm \nu}}} \braket{{\bm \sigma} | \hat{P}^{\bm \theta}_{T|\hat{{\bm \nu}}}}$ from \er{eq:QT} for the case of $\hat{P}_0$ being the flat state. 

As explained above, computing the proposal probability $\hat{P}^{\bm \theta}_T({\bm \sigma})$ cannot be done efficiently using TNs. We can however define an efficient sampling approach that overcomes this problem, as follows. Rather than sampling configurations ${\bm \sigma}$ of the target distribution $P$ consider the problem of sampling initial and final pairs of configurations $({\bm \sigma},{\bm \nu})$ of the noising dynamics, where ${\bm \sigma}$ is sampled from $P$ and  ${\bm \nu}$ is the final configuration after noising for a time $t=T$ having started from ${\bm \sigma}$ at time $t=0$. Their joint probability $P_{0,T}({\bm \sigma},{\bm \nu})$ is given by 
\beq
    P_{0,T}({\bm \sigma},{\bm \nu})
        = 
        \braket{{\bm \nu} | 
            \WW_{t \leftarrow 0} 
            | {\bm \sigma}} P({\bm \sigma}) .
    \label{eq:P0T}
\eeq
Clearly, from $P_{0,T}$ one obtains the target $P$ by marginalisation, $P({\bm \sigma})= \sum_{{\bm \nu}} P_{0,T}({\bm \sigma},{\bm \nu})$, so sampling the former gives access to samples of the latter. Consider similarly the pairs $(\hat{\bm \nu},\hat{\bm \sigma})$ of initial and final configurations for a denoising trajectory from $\hat{t}=0$ to $\hat{t}=T$. The corresponding joint probability $\hat{P}_{0,T}(\hat{\bm \nu},\hat{\bm \sigma})$ is given by 
\begin{align}
    \hat{P}_{0,T}(\hat{\bm \nu},\hat{\bm \sigma}) 
    &= 
    2^{-N} 
    \braket{\hat{\bm \sigma} | 
        \hat{\WW}^{\bm \theta}_{T \leftarrow 0} 
        | \hat{\bm \nu}} 
    \nonumber \\
    & = 
    2^{-N} 
    \frac{P_{\bm \theta}(\hat{\bm \sigma})}{P^{\bm \theta}_T(\hat{\bm \nu})}
    \braket{\hat{\bm \nu} | 
        \WW_{T \leftarrow 0} 
        | \hat{\bm \sigma}} ,
    \label{eq:tP0T}
\end{align}
where we have used that $\hat{\bm \nu}$ is sampled from the flat distribution, and obtained the second equality using \er{eq:tWW}.

In order to sample $P_{0,T}$, an MCMC iteration proposes a move from a current pair of configurations $({\bm \sigma},{\bm \nu})$ to a new pair $(\hat{\bm \sigma},\hat{\bm \nu})$ where $\hat{\bm \nu}$ is sampled uniformly and $\hat{\bm \sigma}$ is obtained by denoising it for $\hat{t}=T$ [most efficiently by directly sampling the MPS \eqref{eq:Qt_update}]. Since the probability being targeted is \er{eq:P0T} and the proposal probability is \er{eq:tP0T}, the corresponding Metropolis acceptance probability reads 
\begin{align}
    \mathcal{A}
    (\hat{\bm \sigma},\hat{\bm \nu} | 
    {\bm \sigma},{\bm \nu}) 
    & = 
    \min
    \left[ 
        1, 
        \frac
        {P_{0,T}(\hat{\bm \sigma},\hat{\bm \nu})}
        {P_{0,T}({\bm \sigma},{\bm \nu})}
        \frac
        {\hat{P}_{0,T}({\bm \nu},{\bm \sigma})}        
        {\hat{P}_{0,T}(\hat{\bm \nu},\hat{\bm \sigma})}
    \right]
    \nonumber \\
    & = 
    \min
    \left[ 
        1, 
        \frac
        {P(\hat{\bm \sigma})}
        {P({\bm \sigma})}
        \frac
        {P_{\bm \theta}({\bm \sigma})}
        {P_{\bm \theta}(\hat{\bm \sigma})}
        \frac
        {P^{\bm \theta}_T(\hat{\bm \nu})}
        {P^{\bm \theta}_T({\bm \nu})}
    \right]
    \label{eq:Metropolis2}
\end{align}
where we have used \era{eq:P0T}{eq:tP0T} in the second line. 

Since the pair $(\hat{\bm \sigma},\hat{\bm \nu})$ is chosen without any direct connection to the previous $({\bm \sigma},{\bm \nu})$ we name this MCMC scheme the {\em disconnected update}. See Fig.~\ref{fig: protocols}(a) for a visualisation: we start with some configuration ${\bm \sigma}$, and use the noising protocol with time $T$ to generate a corrupted sample ${\bm \nu}$; we then draw some new noise sample $\hat{\bm \nu}$ from $\ket{-}$, and apply the denoising protocol to generate the sample $\hat{\bm \sigma}$; the change $({\bm \sigma},{\bm \nu}) \to (\hat{\bm \sigma},\hat{\bm \nu})$ is accepted or rejected using the Metropolis test with acceptance probability \eqref{eq:Metropolis2}. In contrast to \er{eq:Metropolis}, all the quantities appearing in \er{eq:Metropolis2} can be computed efficiently, since $P_{\bm \theta}$ and $P^{\bm \theta}_T$ are represented by MPS, and the ratio ${P(\hat{\bm \sigma})}/{P({\bm \sigma})}$ only involves the known energy $E$, see \er{eq:Boltzmann}.

\subsection{The connected update}
\label{sec: connected}

The second scheme we will consider is one where we directly sample $P$ by proposing changes ${\bm \sigma} \to \hat{\bm \sigma}$, but where the proposal probability for the new configuration depends on the old one. We call this the {\em connected update}, and is depicted in Fig.~\ref{fig: protocols}(b): given the current configuration ${\bm \sigma}$, we will first use the noising protocol for some {\em finite} time $T$ to generate a configuration noised ${\bm \nu}$; then we use ${\bm \nu}$ as the initial configuration for the denoising dynamics for the same extent of time $T$, generating a new sample $\hat{\bm \sigma}$. In contrast to the the disconnected protocol, the noise and denoise branches of the dynamics share the configuration ${\bm \nu}$. In this case the target distribution is $P$ and the probability to propose $\hat{\bm \sigma}$ from ${\bm \sigma}$ is given by
\begin{align}
    \pi(\hat{\bm \sigma}|{\bm \sigma})
    &= 
    \sum_{\bm \nu}
        \braket{\hat{\bm \sigma} | \tilde{\WW}^{\bm \theta}_{T \leftarrow 0} | {\bm \nu}}
        \braket{{\bm \nu} | \WW_{T \leftarrow 0} | {\bm \sigma}}
    \nonumber \\
    &= 
    \sum_{\bm \nu}
        \frac{P_{\bm \theta}({\bm \sigma})}{P^{\bm \theta}_T({\bm \nu})}   
        \braket{{\bm \nu} | \WW_{T \leftarrow 0} | \hat{\bm \sigma}}
        \braket{{\bm \nu} | \WW_{T \leftarrow 0} | {\bm \sigma}} ,
    \label{eq:pi}
\end{align}
where we have used \er{eq:tWW}. The acceptance probability then reads
\begin{align}
    \mathcal{A}
    (\hat{\bm \sigma} | 
    {\bm \sigma}) 
    & = 
    \min
    \left[ 
        1, 
        \frac
        {P(\hat{\bm \sigma})}
        {P({\bm \sigma})}
        \frac
        {\pi({\bm \sigma}|\hat{\bm \sigma})}
        {\pi(\hat{\bm \sigma}|{\bm \sigma})}
    \right]
    \nonumber \\
    & = 
    \min
    \left[ 
        1, 
        \frac
        {P(\hat{\bm \sigma})}
        {P({\bm \sigma})}
        \frac
        {P_{\bm \theta}({\bm \sigma})}
        {P_{\bm \theta}(\hat{\bm \sigma})}
    \right]
    \label{eq:acceptance_correlated}
\end{align}
\Er{eq:acceptance_correlated} is the standard acceptance criterion for Metropolis when attempts are generated from $P_{\bm \theta}$. The acceptance rate will be higher the better  
$P_{\bm \theta}$ approximates $P$, and below we discuss how to learn the parameters ${\bm \theta}$ that define this MPS. Furthermore, the time extent of the noising/denoising $T$ allows us to control how correlated successive proposals are, which can also be optimised in order to better decorrelate successive MCMC samples.

\section{Monte Carlo sampling via DDMs with tensor networks}\label{sec: mcmc}
We now study in detail the integration of MCMC with the update proposals based on sample generation via DDM introduced in the previous section. In particular, we will show the effectiveness of the connected update scheme. For concreteness, in what follows we focus on the problem of sampling the equilibrium state of two specific models, the one-dimensional stochastic Fredkin spin chain \cite{salberger2016fredkin, causer2022slow,morral-yepes2023entanglement}, studied in this section and in the next one, and the two-dimensional Ising model (in a cylindrical geometry) \cite{chandler1987introduction}, studied in Sec.~\ref{sec: learning}. 

Our ultimate aim is to devise an adaptive Monte Carlo method that learns the optimal way to generate proposals for the MCMC using DDMs and TNs. This learning algorithm is presented in Sec.~\ref{sec: learning} below, but we can anticipate some of its features. As the MCMC will be based on proposals from the DDM, cf.\ Sec.~\ref{sec: protocols}, the aim is to learn the optimal initial $P_{\bm \theta}$ and noising time $T$ which together define the denoising dynamics via \er{eq:tWW}. 
The learning algorithm will therefore have three stages: 
(i) Initialisation: there will be a starting guess for $P_{\bm \theta}$, that in general will be far from the target $P$, and a guess for $T$ which will be far from optimal;  
(ii) Improvement: Monte Carlo will filter samples of $P$ which can be used to adapt the parameters of $P_{\bm \theta}$, progressively making it a better approximation to $P$, aided by adjusting $T$ adaptively; and  
(iii) Convergence: learning will eventually reach an optimal $P_{\bm \theta} \approx P$ within its variational class, and the trained MCMC can then be used to generate samples of $P$ efficiently. 
Before defining the learning dynamics in Sec.~VI, in this section we consider each of the three stages separately, exploring the specific aspects of sample proposal/acceptance that affect them. In the next section we assemble these aspects together into an efficient adaptive MCMC scheme. 

In the next subsection we introduce the two models that we study in the rest of the paper. The first of these is the Fredkin spin chain and the second the 2D Ising model. The rest of the section uses the Fredkin chain to consider the separate components of the MCMC scheme of Sec.~VI: the Fredkin chain is convenient to study these issues as is it a spin system with local interactions which, despite being one-dimensional, displays an equilibrium phase transition due to the constrained nature of its configuration space. In this sense, it allows to consider problems relating to the difficulty of sampling across singular changes in the probability of interest. Furthermore, the Fredkin chain is particularly useful as a test case for the methods we introduce as its stationary state can be expressed {\em exactly} as a MPS, and therefore we can benchmark performance of the various components of our Monte Carlo method.

After introducing the models, we use the exact Fredkin MPS to consider sampling with the denoising protocol comparing the disconnected and connected updates of Fig.~\ref{fig: protocols}; this relates to stage (iii) above of sampling once the distribution is learnt. We then investigate the effect of using an approximate MPS to define the sampling, connecting to stage (ii) where only an approximate MPS is used to define the denoising dynamics. The final subsection considers a scenario where this defining distribution is the exact MPS but for a different phase, mimicking the initialisation stage (i) above.

\begin{figure}[t]
    \centering
    \includegraphics[width=\linewidth]{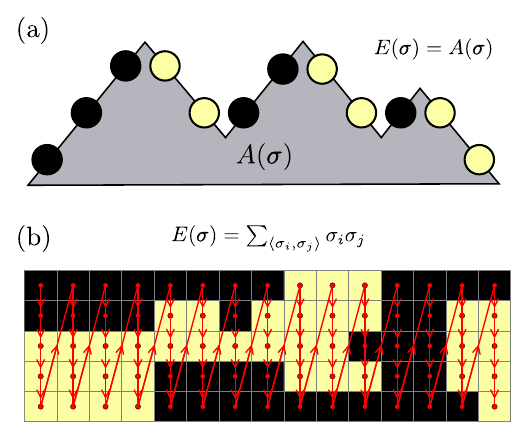}
    \caption{\textbf{Models.} 
    (a) Fredkin spin chain. An example configuration ${\bm \sigma}$ in the half-filling (zero magnetization) sector for size $L=12$, where a black circle is a spin up, and a yellow circle a spin down. The Fredkin constraints state that there must always be at least as many spins up as spins down when counting from the left. The height representation of ${\bm \sigma}$ is obtained by mapping each up spin to a step up, and each down spin to a step down. 
    The energy of the configuration is the area $A({\bm \sigma})$, indicated by the shaded region, under the height field. 
    (b) 2D Ising model. We show a lattice of size $N=15 \times 5$ with PBCs in the vertical direction. A probability vector of this 2D system can be represented by a MPS that ``snakes'' in a 1D path, as shown by the red line.
    }
    \label{fig: models}
\end{figure}

\subsection{Models}
\label{sec: models}

\subsubsection{Fredkin spin chain}
\label{sec: fredkin}

The first system we use to test our methods is the stochastic Fredkin spin chain. Originally introduced as a quantum spin model \cite{salberger2016fredkin}, it was later extended to classical stochastic dynamics \cite{causer2022slow}.
The stochastic Fredkin spin chain \cite{causer2022slow} corresponds to a 1D lattice of $N$ spins $\sigma_{j} = \pm 1$ with $j = 1, \dots, N$, with the conditions on all its configurations that $M_{k} = \sum_{j=1}^{k} \sigma_{j} \geq 0$ for all $k$ and $M_{N} = 0$. This amounts to all allowed configurations having non-negative net magnetisation when counting from left-to-right, and a total magnetisation of zero. 
We denote the set of configurations which obey these criteria by $\mathcal{D}$. This configuration space is equivalent to the set of random walk paths that start in the origin and return to the origin without ever crossing it (or ``excursions'' \cite{majumdar2015effective,rose2021a-reinforcement}) or, equivalently, Dyck paths. The ``physical'' dynamics of the model is defined in terms of exchanges between pairs of neighbouring spins subject to kinetic constraints
that guarantee motion in $\mathcal{D}$. See Ref.~\cite{causer2022slow} for details and references. 

Of interest to us here is the Fredkin equilibrium state \cite{causer2022slow} and not its dynamics. The stationary distribution of the model is defined such that the equilibrium probability for a configuration of spins that obeys all the magnetisation constraints is of Boltzmann form, while for those that do not it is zero. That is, 
\beq
    P({\bm \sigma};\beta) = \frac{1}{Z_\beta}\mathcal{C}({\bm \sigma}) e^{-\beta E({\bm \sigma})} ,
    \label{eq:prob_rwe}
\eeq
where $\mathcal{C}({\bm \sigma})=1$ if ${\bm \sigma}\in \mathcal{D}$ and $\mathcal{C}({\bm \sigma})=0$ if ${\bm \sigma}\notin \mathcal{D}$, $\beta$ plays the role of an inverse temperature (which we will allow to be positive or negative), and the normalisation ${Z_\beta}$ is the partition sum.
The energy function is defined to be ~\cite{causer2022slow}
\beq
    E({\bm \sigma}) = \sum_{j=1}^{N}(N + 1 - j) \sigma_{j} .
    \label{eq:EFredkin}
\eeq
The above energy can be understood in terms of a ``height field'' representation~\cite{causer2022slow}: if each spin represents a step of a random walk, then the position (the height) of the random walker after $j$ steps is $h_{j} = \sum_{k=1}^{j} \sigma_{k}$, see Fig.~\ref{fig: models}. The energy function $E({\bm \sigma})$ can then be understood as the area underneath this height field. In Ref.~\cite{causer2022slow}, it was shown that there is a thermodynamic singularity in 
\er{eq:prob_rwe} at $\beta = 0$ separating three distinct phases: a large area (or ``tilted'') phase for $\beta < 0$, a small area (or ``flat'') phase for $\beta > 0$, and a critical (or ``Coulomb'') phase at $\beta=0$. Furthermore, for any $\beta$ the equilibrium probability vector can be written exactly as an MPS of bond dimension that scales with system size (see Ref.~\cite{causer2022slow} for details).

\subsubsection{2D Ising model on a cylinder}
\label{sec: ising}
As a second example, we will consider Ising model in two dimensions. 
The Ising model describes a lattice of spins, where each spin interacts with its neighbour with energy function 
\beq
    E({\bm \sigma}) = -J\sum_{\langle i,j \rangle} \sigma_{i} \sigma_{j},
    \label{eq:Ising}
\eeq
where $\langle i,j \rangle$ indicates nearest neighbours, and $J$ is a coupling constant which we set to $J = 1$. 
The Ising model in two dimensions undergoes a phase transition in the large size limit at inverse temperature $\beta_{\rm c} = 2 / \log(1 + \sqrt{2})$, from a disordered phase for $\beta < \beta_{\rm c}$ to an ordered phase for $\beta > \beta_{\rm c}$ \cite{Onsager1944crystal,chandler1987introduction}. 

In contrast to the Fredkin chain, for the 2D Ising model we do not have an exact MPS representation of the equilibrium state. While MPS are able to give exact and efficient representations of thermal distributions for one-dimensional systems with short range interactions (effectively representing the transfer matrix of their partition functions), this is not the case in higher dimensions. A way to proceed is to consider a two-dimensional model as a 
long-ranged one dimensional one, in which case there is no guarantee of an efficient MPS representation: for an $L \times L$ lattice, we expect the bond dimension needed to go in principle as $D \sim \mathcal{O}(e^{L})$.

Since we cannot represent the equilibrium probability \er{eq:prob_rwe} with the 2D Ising energy \er{eq:Ising} accurately, in the next section we will demonstrate how the combination of MCMC and DDM generation allows to variationally optimise an MPS that defines an efficient sampling scheme, and which can be considered an alternative approach to standard variational Monte Carlo (see Ref.~\cite{lu2024variational} to see how this is formulated for quantum problems).

We will consider the Ising model in a cylindrical geometry, that is a two-dimensional lattice with dimensions $L_{1} \times L_{2}$ with $L_{2} \geq L_{1}$, and with periodic boundary conditions (PBCs) in the first dimension but not on the second. We define the MPS that approximates probability vectors by ``snaking'' a quasi-1d lattice as sketched in Fig.~\ref{fig: models}(b). In this way  we can investigate the effectiveness of our MPS approach on $d>1$ systems in a controlled manner, while also exploiting the effectiveness of MPS for $d=1$ systems.

\begin{figure}[t]
    \centering
    \includegraphics[width=\linewidth]{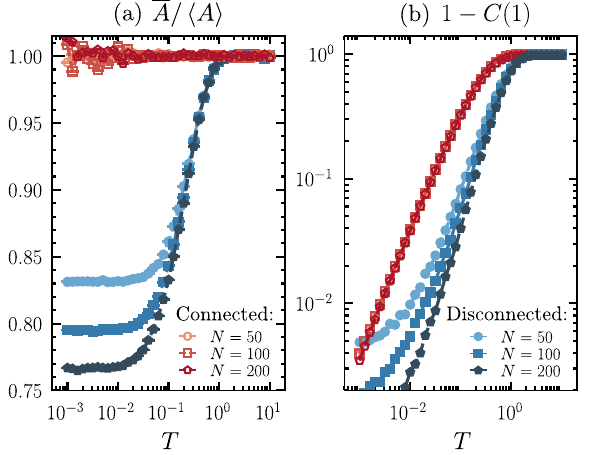}
    \caption{\textbf{Exact sampling of the Fredkin equilibrium distribution.} 
    (a) The area $\overline{A}$ from proposed configurations: for the disconnected scheme (filled/blue symbols) this corresponds to an average over $N_{\rm s} = 10^5$ proposed samples, while for the connected scheme (empty/red symbols) it is an average over $N_{\rm s} = 10^5$ Monte Carlo iterations (as acceptance probability is one). 
    We show sizes $N = 50, 100, 200$ (circles, squares, pentagons, respectively) for \ers{eq:prob_rwe}{eq:EFredkin} at $\beta = 0$.
    (b) Monte Carlo autocorrelation after one iteration, shown as $1 - C(1)$. For both disconnected and connected updates we show results averaged over $N_{\rm s} = 10^5$ MCMC iterations. 
    }
    \label{fig: rwe_exact}
\end{figure}

\subsection{Sampling via denoising from an exact distribution}
\label{sec: sampling_exact}

In the rest of this section we study separate aspects of the sampling using the Fredkin spin chain as an example. 
We first consider the case where the target distribution \eqref{eq:prob_rwe} can be exactly expressed as a known MPS. In an adaptive MCMC scheme like the one we propose in Sec.~VI, this relates to the last stage where the target probability has been fully learnt and one wishes to sample from it, i.e., stage (iii) in the enumeration above.
We consider specifically the case of $\beta=0$ (the critical phase of the Fredkin chain) where the denoising dynamics \er{eq:tWW} is defined from the initial {\em exact} MPS $\ket{P_{\bm \theta}}$ that encodes the equilibrium state \cite{causer2022slow}, so that $\ket{P} = \ket{P_{\bm \theta}}$. 

For the disconnected update, denoising random configurations is guaranteed to propose samples from $P_{\bm \theta}$ for $T \to \infty$. For the connected update, the generated samples are those of $P_{\bm \theta}$ for any $T$. In  
Fig.~\ref{fig: rwe_exact}(a) we illustrate this. For both disconnected and connected schemes we generate $N_{\rm s}$ proposed configurations: for the disconnected case these are $N_{\rm s}$ independent proposals, while for the connected case they are part of a Monte Carlo trajectory of $N_{\rm s}$ iterations, since the acceptance rate for connected updates starting from the exact distribution have acceptance probability one. Figure~\ref{fig: rwe_exact}(a) shows the area averaged over these proposed configurations, $\overline{A} \equiv N_{\rm s}^{-1} \sum_{k=1}^{N_{\rm s}} A({\bm \sigma}^{(k)})$ (in what follows we use over-bar to indicate  empirical mean over samples), relative to the exact equilibrium area. For the connected update, while noisy for small $T$, the estimated areas are compatible with the equilibrium one for all $T$, with fluctuations decreasing with system size. In contrast, for the disconnected update, the average area over the proposed configurations only becomes compatible with equilibrium for long enough $T$, and shows clear size dependence for shorter times. This is an example of the ``mismatch'' characteristic of DMs \cite{de-bortoli2021diffusion}, where the recovery of the initial distribution by denoising is imperfect if $T$ is not long enough.

The results of Fig.~\ref{fig: rwe_exact}(a) show that when starting from $P_{\bm \theta}$ which coincides with target probability $P$ all proposals of the connected update will be accepted irrespective of $T$, while for unconnected updates acceptance will be lower the shorter $T$, in principle only reaching unity in the limit of large $T$. For the more general case where $P$ is unknown and therefore $P_{\bm \theta}$ is only an approximation, we expect the connected update will also lead to higher acceptance of proposed moves, which is one of the desired features of an efficient MCMC simulation. 

A second requirement for efficient MCMC is control on the decorrelation between successive configurations, which is directly related to Monte Carlo convergence. To quantify this we define the normalised autocorrelation function between Monte Carlo samples in the Fredkin chain
\beq
    C(\mu) 
    = 
    \frac{
        \sum_{j=1}^{N}
            \mathbb{E}
            \left[ n_{j}^{(K)}n_{j}^{(K+\mu)} \right] 
            - 
            \mathbb{E}\left[n_{j}\right]^{2}
        }
        {
        \sum_{j=1}^{N}
            \mathbb{E} \left[n_{j}\right] 
            - \mathbb{E}\left[n_{j}\right]^{2}
        }
    \label{eq:MCcorr}
\eeq
where $n_{j}^{(K)} = 2\sigma_{j}^{(K)}-1$ is the occupation at site $j$ at the $K$-th Monte Carlo iteration (including both proposal and acceptance/rejection steps), and $\mathbb{E}[\cdot]$ indicates expectation w.r.t.\ the MCMC. [As for other averages, in practice we estimate $\mathbb{E}[\cdot]$ by the empirical mean over samples $\mathbb{E}[\cdot] \approx N_{\rm s}^{-1} \sum_{k=1}^{N_{\rm s}} (\cdot)$; also, while \er{eq:MCcorr} in principle depends on the iteration $K$, we will only consider it in cases where we start from the exact $P_{\bm \theta}$ so that the MCMC dynamics is stationary, making \er{eq:MCcorr} independent of $K$.]

The Monte Carlo correlator \eqref{eq:MCcorr} starts at $C(\mu=0)=1$ by definition and should go to $C(\mu \to \infty)=0$ if the MCMC is ergodic, being close enough to zero after a finite number of iterations once the MCMC has decorrelated. For simplicity, we focus only on the one-step decorrelation, $C(1)$, which is easiest to compute: it is reasonable to assume that for an ergodic time-inhomogenous Markov process, $C(1) \geq C(\mu) \geq C(1)^\mu$, with equality for the second relation for disconnected updates by definition. 

In Fig.~\ref{fig: rwe_exact}(b) we show the dependence of $1 - C(1)$ on $T$ for both the connected and disconnected schemes. Under the above assumptions, this quantity provides an upper bound on the effective sample size for the connected update, and is the effective sample size for the disconnected update. 
It is interesting to note that despite generating proposals that are independent at every iteration, the disconnected update has a larger autocorrelation
between successive accepted configurations, the larger the shorter $T$, due to a larger rejection rate.

\begin{figure}[t]
    \centering
    \includegraphics[width=\linewidth]{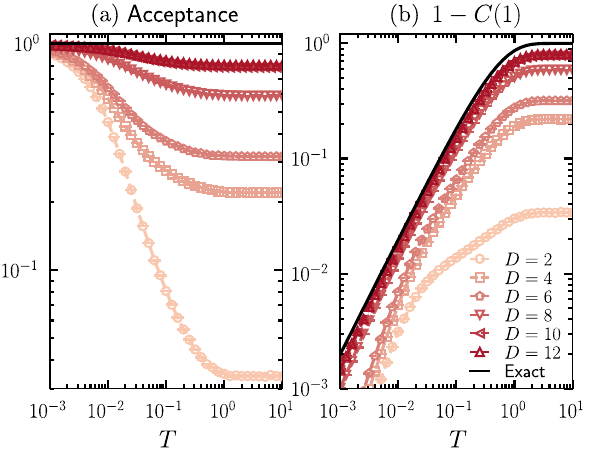}
    \caption{\textbf{Sampling starting from an approximate distribution.} 
    (a) Acceptance probability of the MCMC with the connected update for the Fredkin chain equilibrium state at $\beta = 0$
    as a function of noising/denoising time $T$ for $P_{\bm \theta}$ as an MPS of bond dimension $D$ (symbols), with the sampling from the exact $P$ (and MPS with bond dimension $D=24$)
    for comparison (solid/black line)
    (b) Same by for the one-step Monte Carlo decorrelator $1-C(1)$. 
    Results are for $N = 100$ and $N_{\rm s} = 10^{6}$ samples.
    }
    \label{fig: rwe_truncated}
\end{figure}

\subsection{Sampling via denoising from an approximate distribution}
\label{sec: sampling_approximate}

The next aspect we consider is when the situation where the probability $P_{{\bm \theta}}$ is a only an estimate of the target $P$ and not the exact one as in the previous subsection. This relates to stage (ii) in the adaptive MCMC scheme, where the MPS $P_{{\bm \theta}}$ is progressively improved, but is only an approximation to the target $P$. We can engineer this situation for the Fredkin chain in a controlled manner as we know the exact $P$ in MPS form: we define an approximate MPS $P_{{\bm \theta}}$ and we control the error through its bond dimension $D$ chosen to be smaller than that of the exact state $P$. A natural way to define the best approximate $P_{{\bm \theta}}$ for given $D$ is by minimising its relative entropy, or Kullback-Leibler (KL) divergence, to  the exact $P$,
\beq
    D_{\rm KL}(P \, || \, P_{\bm \theta}) = \sum_{{\bm \sigma} \in \mathcal{D}} P({\bm \sigma}) \log\left( \frac{P(\bm \sigma)}{P_{\bm \theta}(\bm \sigma)} \right) .
    \label{eq:KL}
\eeq
Since the configuration space $\mathcal{D}$ of the Fredkin chain can be sampled exactly from the MPS representation of the equilibrium state at $\beta = 0$ \cite{causer2022slow}, minimising \er{eq:KL} is a tractable problem. For details see App.~\ref{app: ml}.

We then use the MCMC with the connected update to sample the distribution $P$ (at $\beta = 0$) with noising and denoising protocols defined from the approximate MPS $\ket{P_{\bm \theta}} = \ket{\psi_{\bm \theta}} \odot \ket{\psi_{\bm \theta}}$.
Figure~\ref{fig: rwe_truncated}(a) shows the acceptance rate for the updates as a function of the noising/denoising time $T$ for several bond dimensions $D$. In contrast to the case where $P_{\bm \theta}$ coincides with the exact $P$ (full/black line), acceptance probability is less than one and decreases with bond dimension $D$ (i.e., smaller acceptance the less accurate the initial state). Furthermore, the acceptance is larger for smaller $T$ where the proposed updates are smaller.

Despite the smaller acceptance, the more efficient updates are those with larger $T$, as seen from the behaviour of the one-step decorrelator $1 - C(1)$ shown in Fig.~\ref{fig: rwe_truncated}(b) as as function of $T$ for various bond dimensions. This indicates that if the starting state $\ket{P_{{\bm \theta}}}$ is already a reasonable approximation to the target $P$ (for example in the late stages of learning, see below) it is best to use denoising  with large time $T$ which is close to sampling from  $\ket{P_{{\bm \theta}}}$ in an uncorrelated way.

\begin{figure}[t]
    \centering
    \includegraphics[width=\linewidth]{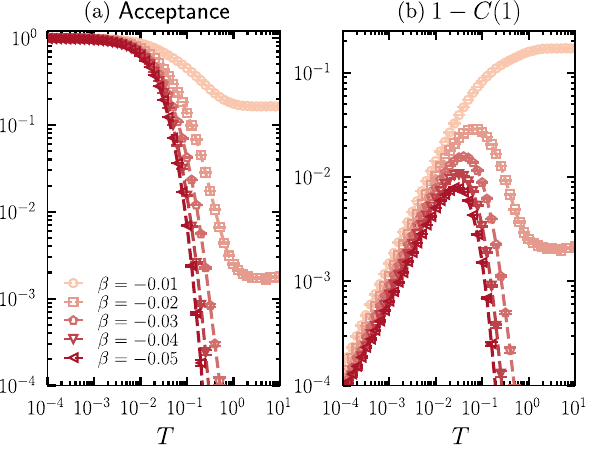}
    \caption{\textbf{Sampling a different target distribution.}
    (a) Acceptance probability as a function of denoising time $T$ for the connected update scheme, where denoising is defined using for 
     $P_{{\bm \theta}}$ the exact Fredkin MPS at $\beta = 0$, and the target $P$ is the Fredkin equilibrium at $\beta \neq 0$, for several $\beta$ in the ``tilted'' phase.  
    (b) One-step Monte Carlo decorrelator as a function of $T$. Results are for $N = 100$ and averaged over $N_{\rm s} = 10^{6}$ MCMC samples.
    }
    \label{fig: rwe_target}
\end{figure}

\subsection{Sampling unknown target distributions}
\label{sec: sampling_target}

Our final preliminary consideration is the case where $P_{\bm \theta}$ is clearly distinct from the target $P$. This relates to the initial stage (i) above. For the case of the Fredkin chain we can engineer this situation by for $P_{\bm \theta}$ using the exact MPS at one value of $\beta$ to generate proposed configurations for another $\beta$. For example, a good test is to use the exact MPS state at $\beta = 0$ to target the distribution at $\beta \neq 0$ which corresponds to a different equilibrium phase \cite{causer2022slow}. 

We first consider the acceptance rate and the one-step decorrelation of the resulting MCMC scheme in its stationary state for the connected update, that is, after enough Monte Carlo iterations have occurred such that the configurations are sampled from the target $P$. In Fig.~\ref{fig: rwe_target}(a) we show that the acceptance probability decreases with increasing $T$, which becomes more pronounced the further the target $\beta$ is from $\beta = 0$ [cf.\ Fig.~\ref{fig: rwe_truncated}(a) where something similar occurs as the approximate MPS deviates from the exact one]. Figure~\ref{fig: rwe_target}(b) in turn shows that the one-step decorrelation is non-monotonic with $T$ [in contrast to Fig.~\ref{fig: rwe_truncated}(b) where there is no drop in decorrelation]. Given that Monte Carlo efficiency requires maximising acceptance and sample decorrelation, the optimal $T$ would be a compromise between the two behaviours seen in Fig.~\ref{fig: rwe_target}. 

\begin{figure}[t]
    \centering
    \includegraphics[width=\linewidth]{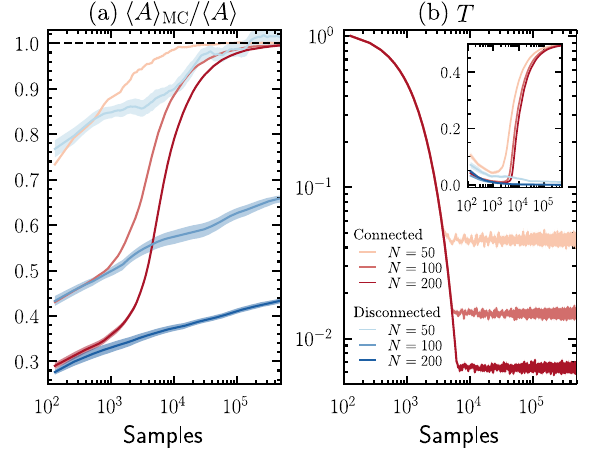}
    \caption{\textbf{Monte Carlo quench and adaptive denoising time.}
    (a) Cumulative average of the area of the Fredkin chain over Monte Carlo samples at $\beta = -0.05$ starting the MCMC from $\beta = 0$, as function of Monte Carlo iterations. The denoising time adapts towards a target acceptance rate of $1/2$. We compare the connected (red) and disconnected (blue) updates, for sizes $N = 50, 100, 200$. The black/dashed line shows the exact equilibrium value. 
    (b) Adaptive denoising time, $T$, as a function of MCMC iterations for the connected update. 
    Inset: Acceptance rate as a function of MCMC iterations for both update schemes. 
    Each experiment is run $10$ independent times, and the curves shown are the average over all experiments. The ribbon shows the standard error.
    }
    \label{fig: rwe_target_quench}
\end{figure}

Secondly, we consider a Monte Carlo ``quench'', that is, the evolution starting from $\beta=0$ of the Monte Carlo dynamics as it equilibrates towards $\beta \neq 0$ (with the denoising protocol is defined from the MPS at $\beta = 0$ as before). We show the corresponding results in Fig.~\ref{fig: rwe_target_quench} for a target $\beta = -0.05$ for various system sizes and for both the connected (red) and disconnected (blue) updates. Anticipating the learning protocol of Sec.~VI, we also use an adaptive scheme which changes the denoising time $T$ based on the acceptance of the last set of samples (set to $128$ in the figure): if the acceptance is greater than $1/2$ then  $T$ is increased, otherwise $T$ is decreased.
In Fig.~\ref{fig: rwe_target_quench}(a) we show the cumulative average of the area, $\braket{A}_{\rm MC}$, as a function of Monte Carlo iterations for both update schemes and for various sizes: while increasing size slows the convergence to the expected value (black dashed line), the connected update (red) is able to reach if faster than the disconnected one (blue).
Figure~\ref{fig: rwe_target_quench}(b) shows the adapted value $T$ for the connected update (initiated at $T = 1$) as a function of Monte Carlo iterations. Note its rapid decrease until it reaches a steady value (which decreases with system size) that allows for efficient sampling at reasonable acceptance rate, here set as the threshold of $1/2$ [the inset to Fig.~\ref{fig: rwe_target_quench}(b) shows convergence to this threshold for the connected update, in contrast to the disconnected update which cannot reach it for the maximum iterations shown].

\section{Learning optimal denoising protocols for Monte Carlo Sampling}\label{sec: learning}
We now have all the elements in place to define an adaptive learning scheme to optimise a denoising protocol for generating Monte Carlo samples. 
The starting point is some a
denoising dynamics defined in terms of 
some initial distribution $\ket{P_{\bm \theta}}$ given by the MPS $\ket{\psi_{\bm \theta}}$, cf.\ Fig.~\ref{fig: mps}, together with a denoising time $T$. In what follows we will only consider the connected update, cf.\ Fig.~\ref{fig: protocols}. 

The Markov chain of the Monte Carlo starts from some configuration ${\bm \sigma}^{(0)}$, which can be sampled directly from $P_{\bm \theta}$. 
If at iteration $k$, the configuration of the Markov chain is ${\bm \sigma}^{(k)}$, it then evolves to configuration ${\bm \sigma}^{(k+1)}$ at the next iteration by accepting or rejecting a proposed configuration generated with the connected update as explained in Sec.~\ref{sec: connected}. In the sampling investigations of Secs.~\ref{sec: sampling_exact}, \ref{sec: sampling_approximate} and \ref{sec: sampling_target}, the MPS $\ket{P_{\bm \theta}}$ that defines the denoising dynamics was kept fixed. Here we devise a scheme by which this MPS is progressively learnt to be a good approximation to the target $P$, together with optimising the denoising time $T$, therefore making the MCMC sampling efficient. 

In order to learn the parameters of the MPS that defines $\ket{P_{\bm \theta}}$ we use as an objective the negative log-likelihood (NLL) of $P_{\bm \theta}$ over $P$,
\beq
    \mathcal{L}[{\bm \theta}] = - \mathop{\mathbb{E}}_{{\bm \sigma} \sim P}
    \Big[ \log P_{\bm \theta}({\bm \sigma}) \Big].
    \label{eq:nll}
\eeq
The parameters ${\bm \theta}$ of the MPS which best approximate the target distribution are found from 
\beq
    {\bm \theta}^{*} = \mathop{\rm argmin}_{{\bm \theta}} \mathcal{L}[{\bm \theta}] . 
    \label{eq:nllmin}
\eeq
The loss \eqref{eq:nll} has a global minimum given by $P_{\bm \theta}=P$, which \er{eq:nllmin} only gives if the variational class spanned by the MPS is large enough. In practice, $P_{\bm \theta}$ can get as close to $P$ as the bond dimension of the MPS allows. A second issue is that in \era{eq:nll}{eq:nllmin} the expectation is taken with respect to the (unknown) target distribution $P$. While we do not have direct access to $P$, we can estimate ${\mathbb E}_P(\cdot)$ in \er{eq:nll} by the empirical average over the samples of $P$ obtained while running the Monte Carlo. 

\begin{figure}[t]
    \centering
    \includegraphics[width=\linewidth]{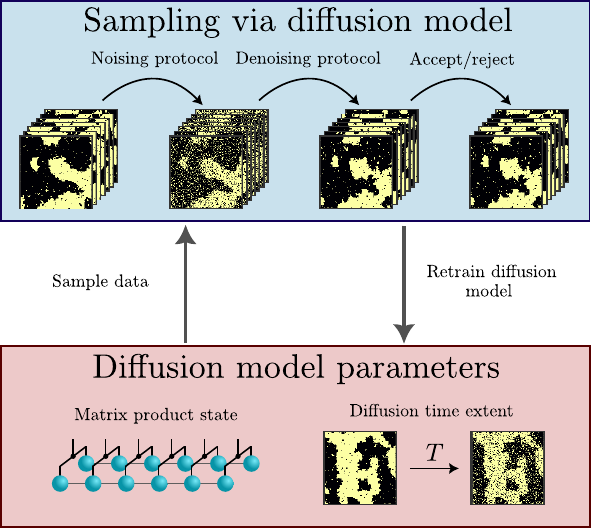}
    \caption{\textbf{Learning optimal denoising protocols.} 
    A sketch of our strategy to learn optimal sampling protocols.
    We start with a batch of $N_{\rm b}$ configures and an untrained MPS which defines the denoising protocol.
    We use the denoising protocol to propose updates the configurations, which are accepted / rejected using the Metropolis criterion.
    The batch of configurations is continuously used to update the MPS using maximum likelihood with stochastic mini-batch gradient descent.
    }
    \label{fig: mcmc}
\end{figure}

Our adaptive MCMC works as follows, see Fig.~\ref{fig: mcmc}:
\begin{enumerate}[label=(\roman*)]
    \item We start from an initial $\ket{P_{\bm \theta}}$ from the untrained MPS $\ket{\psi_{\bm \theta}}$ and with an initial value of $T$. The results from Sec.~V.D suggest that to get good acceptance with a $P_{\bm \theta}$ that is very far from $P$ the initial denoising time $T$ should be small, cf.\ Fig.~\ref{fig: rwe_target_quench}(a).
    However, we will initialise the time to $T = 1$ to demonstrate that the method will decide on a small $T$ without further input. 
    
    \item Rather than a single Monte Carlo trajectory, we run a ``batch'' of $N_{\rm r}$ trajectories in parallel (which we refer to as {\em replicas}). This is required to calculate an empirical average in the loss \eqref{eq:nll}, and has the same computational complexity to running these MCMC trajectories in sequence. The starting configurations for the trajectories in the replica are sampled from the current $\ket{P_{\bm \theta}}$. 
    
    \item From each starting configuration we run the noise-denoise cycle for time $T$ as specified in the connected update scheme of Sec.~\ref{sec: connected}. This results in $N_{\rm b}$ proposed updates, which are accepted or rejected according to \er{eq:acceptance_correlated}. 

    \item In order to learn the optimal MPS we ``weave'' the MCMC iterations and the ``minibatch'' gradient descent minimisation of the loss \eqref{eq:nll}. That is, after each Monte Carlo iteration, we update the MPS $\ket{\psi_{\bm \theta}}$ using an approach akin to that of the so-called density matrix renormalisation group (DMRG), whereby we sweep through each of the local tensors that define the MPS and optimise them according to \era{eq:nll}{eq:nllmin}. Each tensor is optimised by doing one step of gradient descent, where the loss is approximated as the mean over the $N_{\rm b}$ current Monte Carlo configurations. In this manner we progressively minimise the loss in a stochastic fashion. Furthermore, while for a single trajectory proposals with low acceptance rate can give rise to correlated samples, thus increasing the risk of becoming stuck in local minima of the loss (i.e., a form of ``mode collapse''), a benefit of running multiple replicas is that their mutual independence make this problem less likely. For more details on the MPS learning see App.~\ref{app: ml} (and Ref.~\cite{han2018unsupervised} for more general aspects of training generative MPS).

    \item We also adjust the denoising time $T$ using the heuristic approach described in Sec.~\ref{sec: sampling_target}. 
    We aim for a set acceptance probability level (of around $1/2$ in the results below), increasing $T$ if the acceptance over the batch is larger than this target, or reducing $T$ if it is smaller. The results from Secs.~\ref{sec: sampling_approximate} and \ref{sec: sampling_target}, suggest that $T$ should be small in the initial stages of training, cf.\ Fig.~\ref{fig: rwe_target_quench}(a), progressively growing as the MPS gets better, while maintaining a balance between smaller $T$ for larger acceptance and larger $T$ for better decorrelation, cf.\ Fig.~\ref{fig: rwe_truncated}. 
\end{enumerate}
We run this learning scheme until convergence. From then on we can use the trained $\ket{P_{\bm \theta}}$ and $T$ to efficiently sample $P$ using the connected updates, cf.\ Secs.~\ref{sec: sampling_exact} and \ref{sec: sampling_approximate}.

\begin{figure*}[t]
    \centering
    \includegraphics[width=\linewidth]{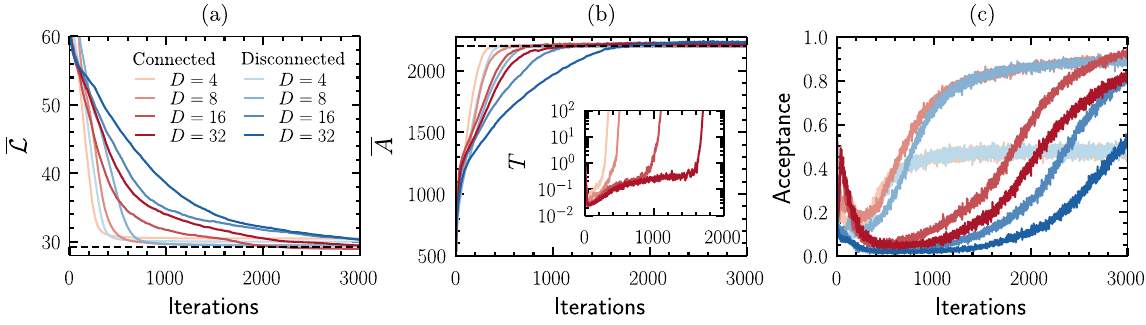}
    \caption{\textbf{Learning optimal DDM sampling for the Fredkin chain.} 
    (a) Mean of the NLL loss \eqref{eq:nll} 
    over a batch of $N_{\rm r} = 1024$ replicas as a function of Monte Carlo iterations, for $\beta = -0.05$ and size $N = 100$. We compare the connected update (red curves) for bond dimensions $D = 4,8,16,32$ to the target NLL (dashed line; estimated from $N_{\rm r}/2$ samples of the exact distribution). We also show for comparison the results obtained with the disconnected update  (blue curves). 
    (b) Batch average of the area as a function of MCMC iterations. Inset: adaptive denoising time as a function of MCMC iterations for the connected update. 
    (a) The average NLL of $512$ test samples, $\braket{\mathcal{L}}$, as a function of sampling iterations.
    The dashed line is the target value.
    (c) Acceptance probability as a function of Monte Carlo iterations for both kind of updates. 
    }
    \label{fig: fredkin_learn}
\end{figure*}

\subsection{Sampling of Fredkin spin chain}

We apply first our adaptive DDM scheme to sample the equilibrium of the Fredkin spin chain at $\beta \neq 0$. 
For the initialisation step (i) we could choose a random MPS of bond dimension $D$. However, since the dimension of the Fredkin subspace $\mathcal{D}$ is polynomially in size smaller than that of unconstrained spin configurations, in order to have a reasonable acceptance initially it is better to start with an MPS that has a good support over $\mathcal{D}$. For the Fredkin chain we can obtain this by choosing $P_{\bm \theta}$ as an MPS of bond dimension $D$ that minimises the distance \eqref{eq:KL} with respect to an empirical distribution of a set of configurations in $\mathcal{D}$ (which for the Fredkin chain can be sampled efficiently from the exact MPS state at $\beta = 0$). The general idea is that for systems with constrained configuration spaces it is sensible to start, if possible, with an untrained MPS but which incorporates partial knowledge of the constraints (if not there will be to be a long initial exploration regime in the training simply to learn the constraints). Furthermore, the overall weight of the MPS on $\mathcal{D}$ is given by 
$\braket{\psi_{\bm \theta} | \hat{\mathcal{C}} | \psi_{\bm \theta}}$,
where $\hat{\mathcal{C}}$ is the projection operator onto $\mathcal{D}$. Since $\hat{\mathcal{C}}$ is an MPO, this weight is easy to calculate, and we use it to determine the level for adjusting the denoising time $T$ in step (v) of our scheme: we increase $T$ if the batch acceptance is larger than $\braket{\psi_{\bm \theta} | \hat{\mathcal{C}} | \psi_{\bm \theta}}/2$, or decrease it otherwise. 

Figure \ref{fig: fredkin_learn} shows various metrics for the learning dynamics as a function of Monte Carlo iterations for the problem of sampling the Fredkin equilibrium at $\beta = -0.05$. 
For the training of the DDM we run $N_{\rm r} = 1024$ replicas. In Fig.~\ref{fig: fredkin_learn}(a) we plot the average loss, cf.\ \er{eq:nll}, for several MPS bond dimensions $D = 4, 8, 16, 32$ (red curves), together with the target value obtained when $P_{\bm \theta}$ is the exact MPS (black dashed line, where the NLL is approximated by averaging over $N_{\rm r}/2$ samples from the exact MPS). For comparison, we also show the results obtained if in step (iii) we use the disconnected update (blue curves): clearly the connected update converges to the target value quicker for all bond dimensions shown. 
Increasing the bond dimension appears to slow the rate of convergence, 
possibly due to the fact that an MPS with a larger bond dimension is able to overfit the fluctuations of the training data, which has a large variance during the training process. On the contrary, MPS with smaller bond dimensions can only learn the most salient features. 
Nevertheless, it is clear that after enough iterations, the MPS with larger bond dimensions converge closer to the exact NLL. 

In Fig.~\ref{fig: fredkin_learn}(b) we show the batch average of the area. We see a similar convergence to the exact value as for the NLL. The inset to Fig.~\ref{fig: fredkin_learn}(b) show how the adaptive denoising time $T$ changes with the learning: is it quickly becomes small at the start of training, cf.\ Sec.~V.D, then increasing and eventually shooting up when the learning has converged (and in the limit $T \to \infty$, connected and disconnected updates coincide). The benefit of using the connected update becomes clear when one considers the acceptance rate, as shown in Fig.~\ref{fig: fredkin_learn}(c): for the larger bond dimensions it is evident that the connected update leads to a larger acceptance rate than the disconnected one.

\begin{figure*}[t]
    \centering
    \includegraphics[width=\linewidth]{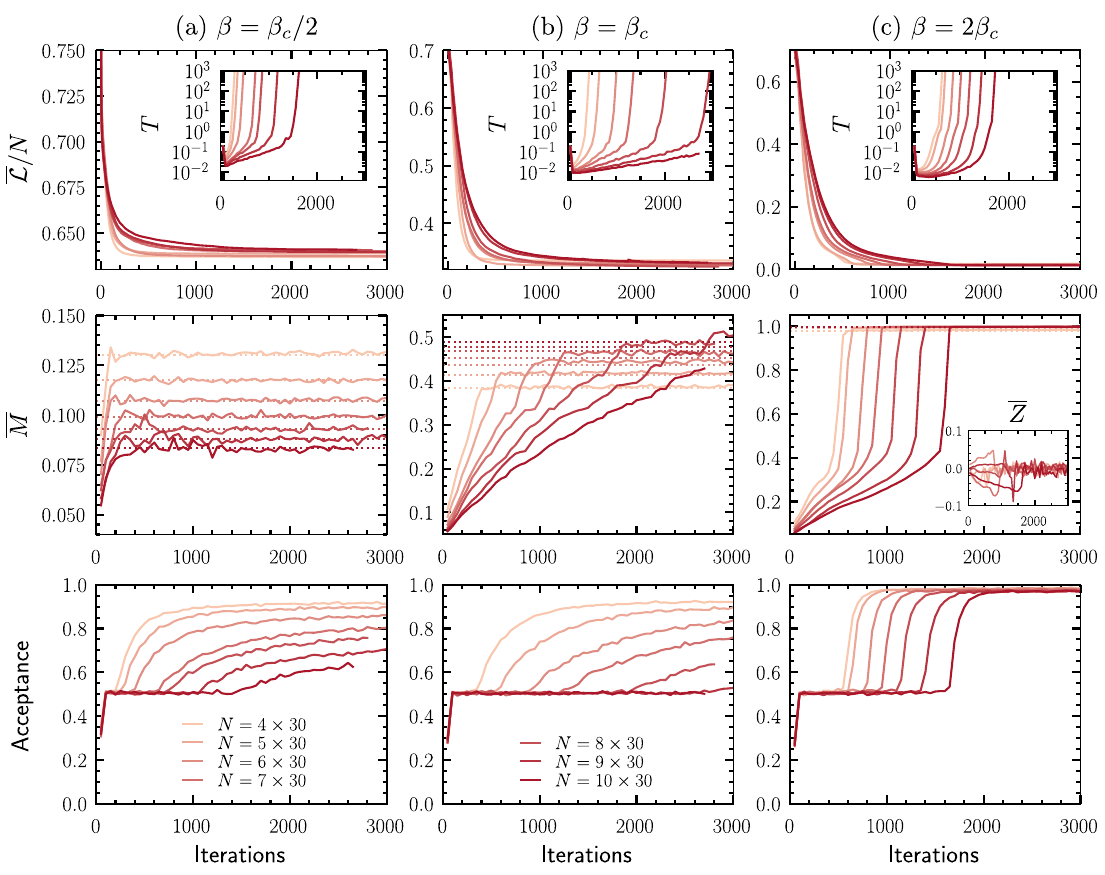}
    \caption{\textbf{Learning optimal DDM sampling for the two-dimensional Ising model.} 
    Column (a): results in the disordered phase, $\beta = \beta_{\rm c}/2$, where $\beta_{\rm c} = \log(1 + \sqrt2) / 2$ is the critical temperature. 
    Top panel: mean of the NLL loss for a batch of $N_{\rm r} = 512$ independent samples obtained from standard Monte Carlo, as a function of Monte Carlo iterations for the MCMC scheme with connected updates, 
    for a range of system sizes with cylindrical boundary conditions. 
    Inset: change in the adaptive denoising time $T$ with MCMC iterations.  
    Middle panel: batch average absolute as a function of training iterations. The dotted lines show the corresponding equilibrium averages. 
    Bottom panel: Acceptance probability as a function of training iterations.
    Column (b): same but at the critical point, $\beta = \beta_{\rm c}$.
    Column (c): same but in the ordered phase, $\beta = 2\beta_{\rm c}$. 
    In all cases shown the maximal bond dimension is $D_{\rm max} = 64$.
    The inset in the middle panel shows the average over spins, averaged over all replicas as a function of MCMC iterations.
    }
    \label{fig: ising_learn}
\end{figure*}

\subsection{Sampling of the two-dimensional Ising model}

We now demonstrate our DDM method for sampling the equilibrium of the two-dimensional Ising model with cylindrical boundary conditions. We saw above, cf.\ Fig.~\ref{fig: fredkin_learn}, that a smaller bond dimension leads to faster learning, even if a larger one gives eventual better results. For this reason we now also adapt value of $D$, as is standard practice in DMRG: we initiate the MPS randomly (as the configuration space of the Ising model is unconstrained) with a bond dimension $D = 1$, and increment the bond dimension every twenty iterations until a maximum of $D_{\rm max}=64$ is reached, thus improving the accuracy of $\ket{P_{\bm \theta}}$.
We also make explicit use of the spatial symmetries of the model: we use the learning scheme with $N_{\rm r} = 512$ replicas, but at each training iteration we spatially reflect in both directions, $L_{2}$, and spatially translated along the depth $L_{1}$ of cylinder for translations $1, \dots, L_{1}-1$, giving a total of $4L_{1}N_{\rm r}$ training samples used in gradient descent.
Note that one could also exploit the spin invariance in \er{eq:Ising} to increase this factor further by two. However, we do not do this here to show that our method is capable of avoiding mode collapse in the ordered phase.

The results are shown in Fig.~\ref{fig: ising_learn} for system sizes $N = L_{1} \times L_{2}$, with $L_{1} = 4$ to $10$ and $L_{2} = 30$.
The panels under column (a) show results for the ordered phase, $\beta = \beta_{\rm c}/2$, column (b) shows results at the critical point, $\beta = \beta_{\rm c}$, and panel (c) in the ordered phase, $\beta = 2\beta_{\rm c}$.
In the top row of Fig.~\ref{fig: ising_learn} is the evolution of the loss for the various system sizes, evaluated with $N_{\rm r}$ independent samples acquired using standard Monte Carlo. In all cases the NLL converges to a steady value. The insets show the adaptive denoising time $T$, where its rapid growth indicates convergence of the training of the DDM (the only exception being for the largest size at $\beta_{\rm c}$), cf.\ Fig.~\ref{fig: fredkin_learn}(b). 
The middle row of Fig.~\ref{fig: ising_learn} shows the absolute value of the total magnetisation averaged over the batch, $\overline{M}$, as a function of Monte Carlo iteration, where $M({\bm \sigma})$ is defined as 
\beq
    M({\bm \sigma}) \equiv
        \left|\frac{1}{N} \sum_{j=1}^{N} \sigma_{j} \right| . 
    \label{eq:M}
\eeq
As for the loss we see convergence to the true average values (dotted lines, computed from standard Monte Carlo for comparison). It is not surprising that convergence is slowest at the critical point where fluctuations are largest. While learning takes longer the larger the system, as one would expect, in no regime it appears to scale exponentially with size. Interestingly, efficient convergence to equilibrium is also fast in the ordered phase. Furthermore, the trained MPS learns about the two ordered phases. This is confirmed by the total magnetisation
\beq
    Z({\bm \sigma}) \equiv
        \frac{1}{N} \sum_{j=1}^{N} \sigma_{j} ,
    \label{eq:Z}
\eeq
shown in the inset to the middle panel of Fig.~\ref{fig: ising_learn}(c): that its value fluctuates around zero, 
when the absolute magnetisation \eqref{eq:M} converges to one, indicates that the learning is able to identify both of ferromagnetic phases, thus avoiding mode collapse in the terminology of machine learning. 

The bottom row of 
Fig.~\ref{fig: ising_learn} shows the Monte Carlo acceptance probability as a function of Monte Carlo iterations. In all cases it quickly stabilises at the $1/2$ level, and then increasing as the learning converges, indicating convergence to the optimal MPS, whose accuracy is limited by the choice of $D_{\rm max}$. Increasing the length $L_{1}$ slows down the increase of the acceptance probability, which is to be expected due to the fixed maximum bond dimension of the MPS. Note that the acceptance is almost one for all systems for $\beta = 2\beta_{\rm c}$, due to the strong correlations in the target distribution.

\section{Discussion}\label{sec: conclusions}
Here we have shown how to efficiently implement discrete diffusion models using tensor networks. By parameterising data, probability vectors and evolution operators with TNs, we showed that a target distribution can be sampled  {\em exactly} without the need to learn stochastic differential equations for the denoising dynamics. As an example application, we used our implementation of DDMs as the proposal generator in a Monte Carlo sampling scheme, with the ability to control the acceptance rate and correlation between proposed moves via properties of the DDM such as the extent of denoising time. 

We also showed how to define an efficient learning scheme where states are approximated within the variational class defined by MPS with a fixed bond dimension. We showed that for Boltzmann sampling where the energy function is known (but not the partition sum), our scheme that combines DDM proposals with Monte Carlo acceptance can learn optimal DDMs to sample the equilibrium distribution effectively. We applied our method to the constrained $d=1$ Fredkin spin chain and the $d=2$ Ising model on a cylinder. In both cases, we found that using DDMs with a learnable MPS and with a ``connected'' denoising dynamics (that overcomes the time-mismatch problem) we could efficiently sample all equilibrium phases, even at (typically hard to sample) coexistence conditions. 

From the technical point of view we can think of several extensions to this work. Here we used the MPS of the initial probability as the learnable quantity, with the subsequent noising/denoising dynamics implemented exactly. One could define a larger variational space where every step of the time evolution is learnable by making the evolution operators into MPOs that can be also trained. A second extension is to implement a similar approach to ours but with TNs in higher dimensions. While MPS are limited to one dimension (or two dimensions on thin cylinders), other TN topologies, such as tree tensor networks (TTNs) \cite{shi2006classical,cheng2019tree} or projected-entangled pair states (PEPS) \cite{verstraete2004renormalization}, could allow for effective applications in in $d>1$. For example, Ref.~\cite{frias-perez2023collective} used PEPS to efficiently sample $d = 2$ distributions using an update which corresponds to the disconnected update presented here.
However, due to the inability to contract them exactly, with PEPS one has to resort to approximate methods which scale poorly with system size. This can lead to a decrease in the acceptance rate as system size is increased. The connected protocol we introduced here might allow for a more efficient sampling with PEPSs more generally.

From the conceptual point of view, there seems to be a fruitful space for considering generative diffusion models from the perspective of statistical physics. Recent examples include studying scaling laws of DMs using mean-field techniques \cite{biroli2023generative}, the description of their generative power in the language of phase transitions \cite{ambrogioni2024the-statistical,biroli2024dynamical}, and their ability to sample equilibrium distributions of disordered models \cite{bae2024a-very}. Since our TNs formulation implements the noising/denoising processes of DDMs {\em exactly}, they could allow for a more comprehensive investigation of these ideas.

The application space of discrete diffusion models is rapidly expanding~\cite{austin2021structured,gruver2023protein} and improvements in representation, sampling, and optimisation could significantly impact many domains. 
Applications focused on lattice models for protein structure and conformational sampling~\cite{robert2021resource-efficient}, which have been presented as a target for quantum algorithms, are a natural future target for our methodology.

\begin{acknowledgments}
    We acknowledge financial support from EPSRC Grant EP/V031201/1.
    LC was supported by an EPSRC Doctoral prize from the University of Nottingham.
    We acknowledge access to the University of Nottingham Augusta HPC service. 
    GMR was supported by the U.S. Department of Energy, Office of Science, Office of Basic Energy Sciences, under Award Number DE-SC0022917
\end{acknowledgments}

\appendix 

\section{Sampling denoising protocols with MPS} \label{app: sampling}
\begin{figure}[t]
    \centering
    \includegraphics[width=\linewidth]{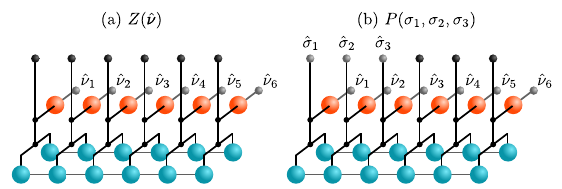}
    \caption{\textbf{Sampling the denoising protocol.}
        (a) The partition function for all configurations $\hat{\bm \sigma}$ is calculated using the same tensor network, but with spins $\hat{\sigma}_{j}$ replaced by the uniform state.
        (b) The marginal probability of observing spins $\hat{\sigma}_{1}$, $\hat{\sigma}_{2}$ and $\hat{\sigma}_{3}$ conditioned on the noise sample $\hat{\bm \nu}$ can be calculated by the tensor network shown.
        The grey spheres indicate spin configurations (as a vector), which are denoted on the diagram, and the black spheres are the uniform distribution for the given lattice site.
    }
    \label{fig: denoising_sampling_appendix}
\end{figure}

The denoising distribution \eqref{eq:Qt_update} for some $\hat{\bm \nu}$ can be sampled exactly sampled as an MPS.
The probability for some configuration $\hat{\bm{\sigma}}$ is given by \er{eq:Qt_update},
\beq
    \hat{P}_{T|\hat{\bm \nu}}^{\bm \theta}(\hat{\bm \sigma}) = P_{\bm{\theta}}(\hat{\bm \sigma}) \braket{\hat{\bm \sigma} | \mathcal{W}_{T\leftarrow 0}^{T} | \hat{\bm \nu}} \frac{1}{P_{T}^{\bm \theta}(\hat{\bm \nu})}.
\eeq
The partition function 
\beq
    Z(\hat{\bm \nu}) = \sum_{\bm \nu} \hat{P}_{T|\hat{\bm \nu}}^{\bm \theta}(\hat{\bm \sigma}) = \braket{- | \hat{P}_{T|\hat{\bm \nu}}^{\bm \theta}}
\eeq
can be calculated explicitly as a TN, which is shown in Fig.~\ref{fig: denoising_sampling_appendix}(a) (with the factor ${1}/{P_{T}^{\bm \theta}(\hat{\bm \nu})}$ not shown).
The black spheres are the local uniform distribution $\ket{-}_{j} = 2^{-1}\sum_{\sigma_{j}} \ket{\sigma_{j}}$ as a tensor.
By contracting this tensor at each lattice site, we sum over the distribution uniformly.

The objective is to sample the configuration $\hat{\bm \sigma} = (\hat{\sigma_{1}}, \dots, \hat{\sigma_{N}})$ from the MPS.
This can be done by considering the marginal of the first $k$ spins,
\beq
    P(\sigma_{1}, \dots, \sigma_{k}) = \frac{1}{Z(\hat{\bm \nu})} \sum_{\j=k+1}^{N} \sum_{\sigma_{j}}  \hat{P}_{T|\hat{\bm \nu}}^{\bm \theta}(\hat{\bm \sigma}).
    \label{eq:app_marginal}
\eeq
The summation in \er{eq:app_marginal} can be calculated by a TN similar to the partition function, see Fig.~\ref{fig: denoising_sampling_appendix}(b).
Notice that the uniform vectors at the sites $j = 1,\dots, k$ are replaced by tensors that represent $\sigma_{j}$.
We then use telescoping to write $\hat{P}_{T|\hat{\bm \nu}}^{\bm \theta}(\hat{\bm \sigma})$ by a product of marginals,
\beq
    \hat{P}_{T|\hat{\bm \nu}}^{\bm \theta}(\hat{\bm \sigma}) =
    \frac{P(\hat{\sigma}_{1})}{Z(\hat{\bm \nu})} \,
    \frac{P(\hat{\sigma}_{1}, \hat{\sigma}_{2})}{P(\hat{\sigma}_{1})} \, \cdots \, 
    \frac{P(\hat{\sigma}_{1}, \dots, \hat{\sigma}_{N})}{P(\hat{\sigma}_{1}, \dots, \hat{\sigma}_{N-1})}.
    \label{product_P}
\eeq
We can exploit \er{product_P} to sample $\hat{P}_{T|\hat{\bm \nu}}^{\bm \theta}(\hat{\bm \sigma})$ using $N$ sampling steps, each given by a fraction in \er{product_P}.

The first is sampling the spin $\hat{\sigma}_{1}$, and the subsequent steps are the distributions of $\hat{\sigma}_{k}$ {\em conditioned} on the previous spins $\hat{\sigma}_{j}$ for $j < k$.
Every factor in \er{product_P} can be calculated exactly as a TN, as shown in Fig.~\ref{fig: denoising_sampling_appendix}.
Contracting such networks can be done with computational cost $\mathcal{O}(ND^{3})$. 
Naively, doing this for each of the $N$ spins gives total sampling cost $\mathcal{O}(N^2 D^{3})$, however, partial contractions can be recycled to give $\mathcal{O}(ND^{3})$, see e.g. Refs.~\cite{schollwock2011the-density-matrix,causer2021optimal}.

\section{Maximum likelihood estimation with MPS} \label{app: ml}
The objective is to optimise the MPS $\ket{\psi_{\bm \theta}}$ such that $\ket{P_{\bm \theta}}$ best approximates $\ket{P}$.
This is achieved by maximising the log-likelihood of the distribution $\ket{P_{\bm \theta}}$ with respect to $\ket{P}$.
That is, we would like to maximise the NLL object function 
\begin{multline}
    \mathcal{L} = \mathop{\mathbb{E}}_{P({\bm \sigma})} \left[\log \frac{\psi_{\bm \theta}^{*}({\bm \sigma}) \psi_{\bm \theta}({\bm \sigma})}{\sum_{{\bm \sigma'}} \psi_{\bm \theta}^{*}({\bm \sigma'}) \psi_{\bm \theta}({\bm \sigma'})} \right]
    \\
    = \mathop{\mathbb{E}}_{P({\bm \sigma})} \left[ \log \frac{\braket{\psi_{\bm \theta} | {\bm \sigma}}\braket{{\bm \sigma} | \psi_{\bm \theta}}}{\braket{\psi_{\bm \theta} | \psi_{\bm \theta}}} \right].
    \label{mps-ll}
\end{multline}
In practice, we will achieve this using gradient descent.
Taking inspiration from standard variational MPS methods \cite{schollwock2011the-density-matrix}, we use a DMRG-like approach.
In this approach, we optimise only a small subset of our variational parameters, namely those of {\em two} neighbouring tensors in the MPS at sites $j$ and $j+1$.
Such parameters are then optimised using gradient descent until the given convergence criteria is met. 
This procedure is iterated across the entire lattice, where we sweep through the MPS from left-to-right and then right-to-left. 
This sweeping process is repeated until convergence.

\subsection{Canonical form}

\begin{figure}[t]
    \centering
    \includegraphics[width=0.6\linewidth]{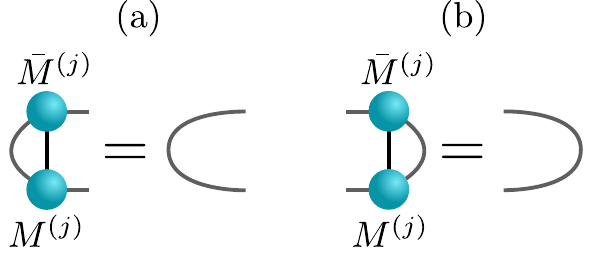}
    \caption{The tensors of the MPS, $M^{(j)}$, can be transformed into an (a) left-canonical representation or (b) a right-canonical representation, described by conditions \ers{left-canonical}{right-canonical}.
    }
    \label{fig: mps_gauge}
\end{figure}

A useful property of MPS is that they have gauge freedom.
Take some invertible $(D, D)$ matrix $X$, and consider the following transformation:
\begin{align}
    M^{(j)} &\to M^{(j)} X, \nonumber
    \\ M^{(j+1)} &\to X^{-1} M^{(j+1)}.
\end{align}
It is obvious that such a transformation leaves the MPS $\ket{\psi_{\bm \theta}}$ unchanged.
This gauge freedom can be exploited to write {\em any} MPS in a so-called canonical representation. 
Consider the tensor $M^{(j)}$.
We say the tensor is {\em left-canonical} or {\em right-canonical} if
\begin{align}
    \sum_{\sigma_{j}} \left(M^{(j)}_{\sigma_{j}}\right)^{\dagger} M^{(j)}_{\sigma_{j}} &= \hat{\mathds{1}},
    \label{left-canonical}
    \\
    \sum_{\sigma_{j}} M^{(j)}_{\sigma_{j}} \left(M^{(j)}_{\sigma_{j}}\right)^{\dagger}  &= \hat{\mathds{1}},
    \label{right-canonical}
\end{align}
respectively.
This is shown as a tensor network in Fig.~\ref{fig: mps_gauge}.
We then say an MPS is in {\em mixed-canonical} representation at site $k$ if $M^{(j)}$ for $j < k$ is in left-canonical representation, and $M^{(j)}$ for $j > k$ is in right-canonical representation.
Any MPS can be written in mixed-canonical representation, which is achieved using singular value decompositions (SVDs), see Ref.~\cite{schollwock2011the-density-matrix} for more details.
Furthermore, the mixed-canonical representation can be easily moved to any other lattice site by use of SVDs \cite{schollwock2011the-density-matrix}.

\subsection{Gradient descent}

\begin{figure}[t]
    \centering
    \includegraphics[width=0.6\linewidth]{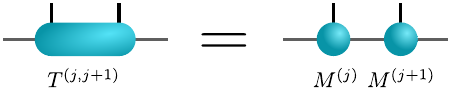}
    \caption{The tensor $T^{(j, j+1)}$ is found by contracting over $M^{(j)}$ and $M^{(j+1)}$.
    }
    \label{fig: mps_contract}
\end{figure}

\begin{figure*}[t]
    \centering
    \includegraphics[width=0.9\linewidth]{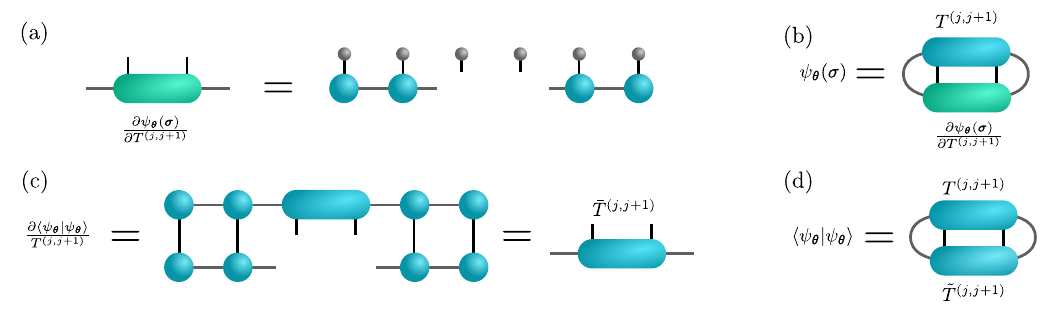}
    \caption{\textbf{Updating the MPS.}
    (a) The value $\psi(\sigma_{1}, \dots, \sigma_{N})$ can be calculated through a TN contraction.
    (b) The gradient of $\psi(\sigma_{1}, \dots, \sigma_{N})$ with respect to the tensor at site $j$, $M^{(j)}$, can be calculated through the partial contraction; contracting all tensors in $\psi(\sigma_{1}, \dots, \sigma_{N})$ except for $M^{(j)}$.
    (c) Calculating the norm $\braket{\psi | \psi}$.
    (d) The gradient of the norm, $\frac{\partial \braket{\psi | \psi}}{\partial M^{(j)}}$ with respect to the tensor $M^{(j)}$.
    }
    \label{fig: mps_updates}
\end{figure*}

The two tensors are contracted to form a single tensor,
\beq
    T^{(j, j+1)} := T^{(j, j+1)}_{\sigma_{j}, \sigma_{j+1}} = M^{(j)}_{\sigma_{j}}M^{(j)}_{\sigma_{j+1}},
\eeq
which is shown diagrammatically in Fig.~\ref{fig: mps_contract}, and is done with computational cost $\mathcal{O}(D^{3}d^{2})$.
We then minimise \er{mps-ll} with respect to the tensor $T^{(j, j+1)}$,
\begin{multline}
    \frac{\partial \mathcal{L}}{\partial T^{(j, j+1)}} =
    \sum_{{\bm \sigma}} \frac{\partial \mathcal{L}}{\partial \psi_{\bm \theta}({\bm \sigma})} \frac{\partial \psi_{\bm \theta}({\bm \sigma})}{\partial T^{(j, j+1)}}
    \\
    = \mathop{\mathbb{E}}_{P({\bm \sigma})} \left[ \frac{1}{\psi_{\bm \theta}({\bm \sigma})} \frac{\partial \psi_{\bm \theta}({\bm \sigma})}{\partial T^{(j, j+1)}}\right] - \frac{1}{\braket{\psi_{\bm \theta} | \psi_{\bm \theta}}} \frac{\partial \braket{\psi_{\bm \theta} | \psi_{\bm \theta}}}{\partial T^{(j, j+1)}}.
    \label{grad}
\end{multline} 
The first term is \er{grad} can be estimated using a mini batch of $b$ samples, ${\bm \sigma}_{k}$ for $k=1,\dots,b$, and the second term can be evaluated exactly, \begin{multline}
    \frac{\partial \mathcal{L}}{\partial T^{(j, j+1)}}
    \approx
     b^{-1} \sum_{k=1}^{b}
    \left[ \frac{1}{\psi_{\bm \theta}({\bm \sigma}_{k})} \frac{\partial \psi_{\bm \theta}({\bm \sigma}_{k})}{\partial T^{(j, j+1)}}\right]
    \\
    - \frac{1}{\braket{\psi_{\bm \theta} | \psi_{\bm \theta}}} \frac{\partial \braket{\psi_{\bm \theta} | \psi_{\bm \theta}}}{\partial T^{(j, j+1)}}.
\label{grad_mb}
\end{multline}
Each gradient in the sum in \er{grad_mb} can be easily calculated as a tensor network calculation, see Fig.~\ref{fig: mps_updates}(a), and has has computational cost $\mathcal{O}(ND^{2})$.
The weights $\psi_{\bm \theta}({\bm \sigma}_{k})$ can be obtained in a similar way.
However, we can recycle the recently contracted gradient to obtain $\psi_{\bm \theta}({\bm \sigma}_{k})$ at an additional cost of $\mathcal{O}(D^{2}$), see Fig.~\ref{fig: mps_updates}(b).

The second term is calculated over in a similar manner.
This time, to calculate the gradient, we must contract over the tensor network shown in Fig.~\ref{fig: mps_updates}(c).
However, note that because we enforce that the MPS is in mixed-canonical form at the site $j$ or $j+1$, it follows that the network can be reduced to the adjoint of the tensor $T^{(j, j+1)}$.
Similarly, $\braket{\psi_{\bm \theta} | \psi_{\bm \theta}}$ is calculated by contracting $T^{(j, j+1)}$ with its adjoint, see Fig.~\ref{fig: mps_updates}(d), and has computational cost $\mathcal{O}(D^{2}d^{2})$.
The total cost of estimating the gradient \er{grad_mb} is $\mathcal{O}(bND^{2} + D^{2}d^{2})$.
Note that while it is possible to just do one iteration of gradient descent before moving onto the next set of tensors, one could also do many iterations.
Here, at each point we do $10$ iterations of GD, each with a learning rate $\alpha$, which yields an effective learning rate of $10\alpha$.
We find that splitting the update into many steps increases the stability of the algorithm.

After the optimisation of the tensor is complete, one must then restore the tensor $T^{(j, j+1)}$ into MPS form, i.e., $T^{(j, j+1)}$ must be decomposed into the tensors $M^{(j)}$ and $M^{(j+1)}$.
This can be done optimally using a singular value decomposition (SVD), $T^{(j, j+1)} = M^{(j)} S M^{(j+1)}$, at computational cost $\mathcal{O}(D^{3}d^{3})$.
Note that doing this procedure will result in a new bond dimension with $D' = dD$.
In practice, to avoid the bond dimension growing exponentially, one must use a truncated decomposition, which keeps only the $D$ largest singular values, and discards the rest.
Finally, the diagonal matrix of singular values $S$ can be contracted into either $M^{(j)}$ or $M^{(j+1)}$.
Doing so will restore the MPS into mixed canonical form at site $j$ or $j+1$ respectively, and thus which tensor it is contracted into is decided by the direction which we are sweeping over.

\end{document}